\documentclass[11pt]{article}
\usepackage{a4wide}
\usepackage{amsfonts,amsmath,amssymb}
\usepackage{psfrag}
\usepackage{graphics}
\usepackage{subfigure}
\usepackage{ifpdf}
\usepackage{epsf}
\usepackage{color}
\ifpdf
\usepackage{epstopdf}
\usepackage{hyperref}
\else
\usepackage[hypertex]{hyperref}
\fi

\newcommand{\nn}[2]{#1}

\newcommand{\eref}[1]{(\ref{#1})}

\newcommand{\blank}[1]{}

\newcommand\sect[1]{\section{#1}\setcounter{equation}0} 

\newcommand\void[1]       {}

\newcommand\be            {\begin{equation}}
\newcommand\bea           {\begin{eqnarray}}
\newcommand\rd             {{\mathrm d}}
\newcommand\ee            {\end{equation}}
\newcommand\eea           {\end{eqnarray}}

\newcommand\Mc            {\mathcal{M}}

\newcommand\Pc            {\mathcal{P}}

\renewcommand\vec[1]{{\vert{#1}\rangle}}

\newcommand\cev[1]{{\langle{#1}\vert}}

\newcommand\vac{{\vec 0}}

\def\3pt#1#2#3{{\langle{#1}|{#2}|{#3}\rangle}}

\def\cbI#1#2#3#4#5#6#7#8{
\setlength{\unitlength}{#1sp}%
\hbox to 2.8cm{\raise -2mm
\vbox{
\begin{picture}(1990,757)(4600,-2483)
\thicklines
{\put(4850,-1860){\line( 1,0){800}}}%
{\put(4850,-2460){\line( 1,0){800}}}%
{\put(5250,-1860){\line( 0,-1){600}}}%
\put(4775,-2460){\makebox(0,0)[rc]{$#2$}}
\put(4775,-1860){\makebox(0,0)[rc]{$#3$}}
\put(5750,-1860){\makebox(0,0)[lc]{$#4$}}
\put(5750,-2460){\makebox(0,0)[lc]{$#5$}}
\put(5300,-2160){\makebox(0,0)[lc]{$#6$}}
\end{picture}}}%
}

\def\cbb#1#2#3#4#5#6#7#8{
\setlength{\unitlength}{#1sp}%
\hbox to 3.6cm{\raise -2mm
\vbox{
\begin{picture}(2700,800)(4301,-2463)
\thicklines
{\put(4800,-1960){\line( 0,-1){500}}}%
{\put(5700,-1960){\line( 0,-1){500}}}%
{\put(4250,-2460){\line( 1, 0){2000}}}%
\put(4200,-2460){\makebox(0,0)[rc]{$#2$}}
\put(4800,-1900){\makebox(0,0)[cb]{$#3$}}
\put(4800,-2560){\makebox(0,0)[ct]{$#7$}}
\put(5250,-2380){\makebox(0,0)[cb]{$#4$}}
\put(5700,-1900){\makebox(0,0)[cb]{$#5$}}
\put(5700,-2560){\makebox(0,0)[ct]{$#8$}}
\put(6350,-2460){\makebox(0,0)[lc]{$#6$}}
\end{picture}}%
}}

\def\thefootnote{\fnsymbol{footnote}}
\begin{document}

\begin{flushright}  {~} \\[-12mm]
{\sf KCL-MTH-11-10}\\[1mm]
\end{flushright} 

\thispagestyle{empty}

\begin{center} \vskip 14mm
{\Large\bf   The renormalisation group for the Truncated
  Conformal Space Approach on the cylinder}\\[20mm] 
{\large 
P.~Giokas~~\footnote{Email: philip.giokas@kcl.ac.uk} and G.M.T.~Watts~~\footnote{Email: gerard.watts@kcl.ac.uk}
}
\\[8mm]
Department of Mathematics, King's College London,\\
Strand, London WC2R 2LS -- UK

\vskip 22mm
\end{center}

\begin{quote}{\bf Abstract}\\[1mm]
In this paper we continue the study of the truncated conformal space
approach to perturbed conformal field theories, this time applied to
bulk perturbations and focusing on the
leading truncation-dependent corrections to the spectrum. 
We find expressions for the leading terms in the ground state
energy divergence, the coupling constant renormalisation and the
energy rescaling. We apply these methods to problems treated in two
seminal papers and show how these RG improvements greatly increase the
predictive power of the TCSA approach. 
One important outcome is that the TCSA spectrum of excitations is
predicted not to converge for perturbations of conformal weight
greater than 3/4, but the ratios of excitation energies should converge. 
% We illustrate this for both a relevant and an irrelevant perturbation. 

\end{quote}

%\tableofcontents

\vfill
\newpage 

\setcounter{footnote}{0}
\def\thefootnote{\arabic{footnote}}

\sect{Introduction}

The Truncated Conformal Space approach (TCSA) of Yurov and
Zamolodchikov \cite{YZ1} has been a widely-used method
to study the finite-size dependence of perturbed two-dimensional
conformal field theories for quite some time
\cite{LCM1,KM1,BPTW,PT06,MM08,MT09,LTD09}. It is based on truncating 
the infinite 
dimensional Hilbert space to a finite-dimensional system on which the
Hamiltonian is studied numerically.\footnote{%
In this paper we study the original form due to Yurov and
Zamolodchikov, not the revised version of
\cite{K1,K2}. 
}%. 
It has been known for a long time
that the method has various convergence problems which can reduce its
effectiveness \cite{LCM1,KM1}.
The principal problems that have been noted before are the divergence
of, and 
differences between ground state contributions in different
sectors. 
Once these have been taken into account by considering only
differences of energy levels in the same sector, the effects of
truncation can still be important.
There has been some interesting work on extrapolation in truncation
level \cite{BPTW,PT06}, assuming an underlying but unknown scaling
behaviour. 
Here, 
we conjecture that the most important corrections after the ground
state divergence are 
a renormalisation of the
coupling constant and a renormalisation of the energy scale. 
Our main results are perturbative expressions for the leading
coupling-constant and energy-scale renormalisations. We apply these to
the tri-critical Ising model as considered in both \cite{LCM1} and
\cite{KM1} and show that the behaviour of the TCSA results is greatly
improved. 
We also find numerical results for the renormalisation and
rescaling for the tri-critical Ising model and show that the TCSA
approach remains very accurate even when these two effects become large.

One important result we find is that while the TCSA estimates of
energy gaps converge for perturbations with conformal weight less than
$3/4$, they do not for those with weight greater than $3/4$. We show
this in the case of the minimal model $M_{9,10}$ perturbed by a field
of weight $4/5$. This failure of convergence is entirely due to the
divergence of the energy rescaling; once the energy rescaling is taken
into account the TCSA estimates converge as well as before. In other
words, the TCSA estimates of ratios of energy gaps converge even if
the gaps themselves do not. 

The paper is organised as follows.
We first introduce the TCSA approximation and present the problems to
be addressed in section \ref{sec:intro}. We then discuss the perturbative
results for the coupling-constant and energy renormalisations in
sections \ref{sec:cc} and \ref{sec:er}, 
and present the results for the models considered in \cite{LCM1} and
\cite{KM1} in sections \ref{sec:lcm} and \ref{sec:km}. 
In section \ref{sec:div} we then discuss the problems occurring with
$h>3/4$ and in section \ref{sec:strip} the complications for perturbed
models on the strip.
%  and finally in section \ref{sec:ir} we  comment on perturbations by
%  irrelevant fields. In this case, both the energy rescaling and the
%  coupling-constant renormalisation diverge, but the TCSA still seems
%  to provide accurate results once these renormalisations are taken
%  into account, in particular we present TCSA evidence in support of
%  the identification of the irrelevant perturbations of the minimal
%  models in \cite{FQR1}.  

\sect{The TCSA approach and its errors}
\label{sec:intro}

\subsection{The TCSA approach to bulk perturbations}

At present, the TCSA can be applied to bulk perturbations in two
different arenas: when the model is defined on a circle and when it is
defined on a strip. The original paper of Yurov and Zamolodchikov \cite{YZ1}
treats the cylinder case, the boundary case was initiated in
\cite{Dorey:1997yg}. 
We consider here  the case
of the cylinder, and defer the strip to section \ref{sec:strip}.

We start with a CFT defined on a cylinder of circumference $R$ which we
take to be a strip $0\leq y<R$ of width $R$ in the complex plane
with coordinate $z=x+iy$, and with the two edges of this strip
identified.
For the details of CFT, see \cite{YBk}.

The unperturbed Hamiltonian generating translations along the cylinder
is 
\be
 H = \int_0^R T_{xx}\;\frac{\rd y}{2\pi}
\;.
\label{eq:HcftA}
\ee
We will map the strip to the complex plane with coordinate
$w=\exp(2\pi z/R)$ in terms of which the CFT Hamiltonian is 
\be
 H = \frac{2\pi}{R}\left( L_0 + \bar L_0 - \frac{c}{12} \right)
\;,
\label{eq:Hcft2A}
\ee
where $L_0$ and $\bar L_0$ are the zero modes of the two copies of the
Virasoro algebra present in the theory defined on a plane.

We are interested in perturbations by one or more bulk fields
$\varphi_i(x,y)$. We take
these to be spinless quasi-primary fields of equal left and right conformal
dimensions $(h_i,h_i)$. 
If the coupling to these fields are $\mu_i$ then the perturbation is
given by an addition to the action 
\be
\delta S = \int \sum_i \mu_i \varphi_i(x)\,\rd^2 x
\;.
\ee
When mapped to the upper half plane this gives the perturbation to the 
Hamiltonian as
\be
 \delta H = 
 \sum_i\mu_i\left(\frac{R}{2\pi}\right)^{1-2h_i}\,
   \int_{\theta=0}^{2\pi}\varphi_i(e^{i\theta})\,\rd\theta
\;,
\label{eq:Hpcft1A}
\ee
where $w = r\exp(i\theta)$ so that $y = (2\pi\,\theta/R)$.

Since the circle has rotational symmetry, one can restrict attention
to the rotationally invariant states on which $L_0-\bar L_0 = 0$.
On these states, we can perform the $\theta$--integral in
\eref{eq:Hpcft1A}
so that the perturbed Hamiltonian becomes
\be
 H = \frac{2\pi}{R}\left[
 L_0 + \bar L_0 - \tfrac{c}{12} 
+ 
 \sum_i(2\pi)^{1-y_i} \lambda_i \,
   \varphi_i(1).
  \right]
\;,
\label{eq:Hpcft1B}
\ee
where $y_i = 2 - 2h_i$ and $\lambda_i = \mu_i R^{y_i}$ are
dimensionless coupling constants. This is the TCSA Hamiltonian of \cite{YZ1}.

The TCSA method is to truncate the Hilbert space in some prescribed
manner and take the Hamiltonian to be \eref{eq:Hpcft1B} on the
truncated space. There are at least two ways this has been
implemented.

\subsubsection{Truncation by level}
 
The first method is to truncate the space to level $n$ in each representation
in the Hilbert space, as in \cite{YZ1}. This means that the maximum
value of $L_0+\bar L_0$ in each representation is $(\nn{2n}{n}+h_i+\bar h_i)$.
This can cause problems if the Hilbert space includes some
representations with large values of $h$, as these high level states
cause distinct qualitative changes to the spectrum.

\subsubsection{Truncation by total energy}

The second method is to truncate so that $L_0+\bar L_0 \leq \nn{2n}{n}$; this
means that the maximum value of $L_0+\bar L_0$ in each representation
is $\nn{2n}{n}$, and so if a representation has $h>n$, it will be completed
excluded from the TCSA space at level $n$. This method has the
advantage that high weight representations do not unduly affect the
TCSA space, that the space at level $n$ is smaller than in the
truncation by level, but has the disadvantage that the truncation
affects are harder to deal with analytically.

These two methods will give very similar results when $n \gg h$ for
all $h$ in a particular model.
It is worth pointing out that choosing a different truncation method
can have dramatic effects -- the ``mode truncation'' investigated by
T\'oth in \cite{GZsT} has a very different behaviour.
We shall use the level truncation unless stated otherwise.

The operator $H$ in \eref{eq:Hpcft1B} is dimensionful, and it is far
preferable to work with dimensionless operators, or dimensionless
eigenvalues. In the case of bulk perturbations, there are two natural choices. 

\subsubsection{Flows ending in massless theories}

If the IR limit of the flow is a massless theory, then it is natural
to work with the scaling functions defined in terms of the energy
eigenvalues $E_n$ as 
\be
 e_n(\lambda_i) = \left(\frac{R}{2\pi}\right)E_n(\lambda_i)
\;,
\ee
% where $\lambda_i = \mu_i R^{y_i}$ are dimensionless coupling
% constants. 
The scaling functions are expected to flow to the 
eigenvalues of the operator
\be
 \left( L_0 + \bar L_0 - \tfrac{c}{12}  \right)_{IR}
\;,
\ee
at the IR fixed point with a correspondingly simple spectrum.

\subsubsection{flows ending in massive theories}

If the IR limit of the flow is a massive theory, then the mass of the
lightest stable particle, $m$, gives a natural scale and one considers
the dimensionless operator
\be
 \frac{H}{m}
= 
 \frac{2\pi}{r}\left[
 L_0 + \bar L_0 - \tfrac{c}{12} 
+ 
 \sum_i(2\pi)^{1-y_i} \frac{\lambda_i}{m^{y_i}} r^{y_i}\,
   \varphi_i(1).
  \right]
\;,
\label{eq:HpcftA2}
\ee
where $r = mR$ is a dimensionless variable and $\lambda_i m^{-y_i}$
are a set of numbers.

\subsubsection{Problems with the TCSA}

The original model studied with the TCSA by Yurov and Zamolodchikov
was the Lee-Yang model perturbed by its primary field
$\varphi_{(1,3)}$ of weight $h=-1/5$. The results in this model,
whether as in the original case on the cylinder \cite{YZ1}, or on the
strip \cite{Dorey:1997yg}, have been exceptionally accurate, up to
14 digits for some quantities. When the TCSA method was applied more
generally in 
\cite{LCM1} and \cite{KM1}, it was clear that this was not always the
case. There were several problems identified in~\cite{LCM1} for the
perturbation of the tri-critical Ising model by the primary field of
weight $3/5$; we
discuss these problems in turn.

The first
is that the Hilbert space of the model may split up into several
sectors and the TCSA eigenvalues in these sectors may differ by
unphysical amounts. 
In our prototypical examples of perturbations by $\varphi_{13}$ and
$\varphi_{31}$, the perturbation will only couple together
representations in the same row or same column of the Kac table.
For example, the tri-critical Ising model has representations
$\{(r,s); 1\leq r\leq3, 1\leq s\leq4, r+s\,\mathrm{even}\}$. Under the
action of
$\varphi_{(1,3)}$ these fall into three sectors which we denote
$(r,*)$ for $r=1,2$ and $3$.
Under the action of
$\varphi_{(3,1)}$, they split into four sectors,  which we denote
$(*,r)$ for $r=1,2,3$ and $4$.
Under the massless flow generated by $\varphi_{(1,3)}$, the $(*,r)$
sector flows into the $(1,r)$ representation of the Ising model. 
Under the massive flow generated by $\varphi_{(1,3)}$, the 
three sectors flow into three sectors of the massive model,
corresponding to the splitting of the three degenerate ground states
of the massive kink model.  
Under the irrelevant flow generated by $\varphi_{(3,1)}$, the
$(*,r)$ sector flows into the corresponding $(r,*)$ sector of the
minimal model $M(5,6)$. 
%We illustrate this in figure \ref{fig:ir1}
%where the $(*,1)=(1,1)\oplus(3,1)$ sector flows into the
%$(1,*)=(1,1)\oplus(1,3)\oplus(1,5)$ sector.
The problem is that the TCSA gives slightly different results for the
ground state energy is the different sectors, so that it may be
impossible to get accurate results for energy differences of
states in different sectors. Sometimes interpolation in truncation
level or size of the truncated space produces sensible results, as was
reported in \cite{KRW}.
The main result we have to report is that the energy rescaling formula
we find in section \ref{sec:er} is slightly different in different
representations, and that this effect is most noticeable for the
ground states in each sector. This greatly improves
the difference between the ground states in the different sectors, as
can be seen in figure \ref{fig:kmp12b}.

Secondly, if the weight of the perturbing field
becomes larger than 1/2, the conformal perturbation expansion and
correspondingly the TCSA eigenvalues, become divergent. This has the
result that the ground state energy of the TCSA system does not
converge with increasing truncation level. 
This was observed in \cite{LCM1} and discussed in
\cite{KMB370}. 
It was realised that this divergence can arise purely from the second order
contribution for which there is an exact expression. Subtracting the
divergent part of this expression then gives a revised TCSA estimate
which will converge, with increasing truncation level, to the
perturbed conformal field theory result, as observed by Tak\'acs
\cite{gabor}. We re-derive this leading term as part of our treatment of
the energy rescaling in section \ref{sec:er}.
We illustrate the effectiveness of these subtractions in the case of the
massless perturbation of the 
tri-critical Ising model considered in \cite{LCM1}.
We show in figure \ref{fig:cp2} that this works well in the case of
the tri-critical Ising model perturbation considered problematic in
\cite{LCM1} - after subtraction of the leading divergence, the ground
state energy does then converge for this perturbation.

Finally, it also appears
that the ``scaling region'' is not easily reached, the region where
the eigenvalues scale with truncation level in the expected manner.
Our solution is that suggested in \cite{LCM1},
a careful consideration of the scaling of the model with system size
which we show reduces, in the cases considered in \cite{LCM1}, to a
renormalisation of the coupling constant and a
representation-dependent re-scaling of the  
Hamiltonian. This is almost the same as we found in \cite{W2011}.
As in \cite{W2011}, perturbative expansions for these effects can be
found by considering the change in the energy eigenvalues with
truncation level. We consider first the coupling constant
renormalisation in section \ref{sec:cc}, and then the ground state
energy and energy rescaling in section \ref{sec:er} and apply them to
the tri-critical Ising model in section \ref{sec:tcim}.

\sect{Coupling constant renormalisation}
\label{sec:cc}

The derivation of the coupling constant renormalisation is a
straightforward generalisation of the boundary case.  
We assume that the perturbed correlation functions on the cylinder
with coordinates $(x,y)$ are given by the insertion of the expression
\be
  \Pc \exp \left(
  - \sum_i 
    \mu_i 
    \int_{x=-\infty}^{\infty} 
    \int_{y=0}^{R}
    P_n\, \varphi_i(x,y)_{cyl} \,P_n\;\rd y \rd x
  \right)
\;,
\ee
in the unperturbed expressions, where $\Pc$ denotes path ordering and
$P_n$ denotes the projector onto states at level $n$ or lower.
After mapping to the plane with $w=exp(2\pi z/R)$, $r=\exp(2\pi x/R)$,
$\theta=2\pi y/R$, this becomes
\bea
&&
  \Pc \exp\left(
  - \sum_i 
    \lambda_i (2\pi)^{-y_i}\int_{r=0}^{\infty} 
    \int_{\theta=0}^{2\pi}
    P_n\, \varphi_i(w,\bar w)_{cyl.} \,P_n\;
    \frac{r \rd r\rd \theta}{r^{y_i}}
  \right)
\nonumber\\
&=&
 1 
  - \sum_i 
    \tilde\lambda_i 
    \int_{r=0}^{\infty} 
    \int_{\theta=0}^{2\pi}
    P_n\,  \varphi_i(w,\bar w)_{cyl.}  \,P_n\;
    \frac{\rd r\rd \theta}{r^{y_i-1}}
  + \ldots
\;,
\label{eq:cc1}
\eea
where $y_i = 2 - 2 h_i$ and $\tilde \lambda_i = (2\pi)^{-y_i}\lambda_i$
is introduced for convenience.
We require that the perturbed correlation functions be invariant when
the truncation level $n$ is changed. The simplest way to find the
leading order change in the coupling constants is to consider the
matrix elements of the
integrand of $\int_0^\infty r^{1-y}\rd r$ in
\eref{eq:cc1} taken at $r=1$ and taken in 
the states $\cev{\varphi_i}\ldots\vac$, that is we consider 
\be
  Z_{i,n}
= - (2\pi)\tilde\lambda_i
 + \sum_{j,k}(2\pi)\tilde\lambda_j\tilde\lambda_k
   \int_{r=0}^1 \int_{\theta=0}^{2\pi}
   \cev{\varphi_i}\varphi_j(1,1) P_n \varphi_k(r,\theta)\vac
   \frac{\rd r\rd \theta}{r^{y_k-1}}
 + \ldots
\;,
\ee
where we have performed one of the angular integrations. We have also
used the first of the properties of the primary fields
\bea
  \cev{\varphi_i} \varphi_j(w,\bar w)\vac 
&=&
  \delta_{ij} |w|^{-2h_i}
\;,
\nonumber
\\
  \cev{\varphi_i} \varphi_j(1,1)\varphi_k(w,\bar w)\vac
&=& \frac{C_{ijk}}{|1 - w|^{2(h_i-h_j-h_k)}}
\label{eq:defs}
\eea
Requiring $Z_{i,n}=Z_{i,n-1}$, we find
\be
 \tilde\lambda_i(n) - \tilde\lambda_i(n-1)
 = \sum_{j,k} 
   \tilde\lambda_j \tilde\lambda_k
   \int_{r=0}^1\int_{\theta=0}^{2\pi}
   \cev{\varphi_i} \varphi_j(1,1) \left[ P_{n}-P_{n-1}\right]
   \varphi_k(r,\theta)\,\vac
   \frac{\rd r\rd \theta}{r^{y_k-1} }
\;.
\label{eq:cc2}
\ee
From \eref{eq:defs}, 
\be
   \int_0^{2\pi}
   \cev{\varphi_i} \varphi_j(1,1) \left[ P_{n}-P_{n-1}\right]
   \varphi_k(r,\theta)\,\vac
   \rd\theta
= 
  2\pi C_{ijk} \left[
   \frac{\Gamma( h_j+h_k-h_i+\nn{n}{n/2} )}
    {\Gamma( h_j+h_k-h_i )\Gamma(\nn{n}{n/2}+1)}
  \right]^2
  \, r^{2n}
\;.
\label{eq:cc3}
\ee
Substituting \eref{eq:cc3} in \eref{eq:cc2}, performing the $r$
integral and expanding out to leading order in $n$, we get
\be
  n \frac{\rd\tilde\lambda_i}{\rd n}
  \simeq 
  n (\tilde\lambda_i(n)-\tilde\lambda_i(n-1))
  \simeq
  \sum_{jk}
  n^{y_i-y_j-y_k} 
  \tilde\lambda_j
  \tilde\lambda_k
  \frac{2\pi C_{ijk}}{\Gamma(h_j+h_k-h_i)^2}
\;.
\ee
As we see, there are corrections to $\lambda_i$ from all pairs of
fields $\varphi_j$, $\varphi_k$ which couple to $\varphi_i$, but that
those for which $h_i-h_j-h_k>0$, i.e. those which appear in the regular
part of the operator product expansion, do not give important corrections.
In the simplest case where we consider the perturbation by a single
field where the only primary fields occurring in the singular part of
its operator product expansion are the identity and the field itself,
this gives for $\lambda(n)$,
\be
  n \frac{\rd\lambda}{\rd n}
  =
  \frac{\lambda^2}{(2\pi n)^y}
  \frac{2\pi C}{\Gamma(h)^2}
  + O(\lambda^3)
\;,
\label{eq:rgde}
\ee
where $C$ is the three-point coupling. 
If $y>0$, this can be integrated to find the effective ``exact'' coupling
$\lambda_\infty$ in terms of the TCSA coupling $\lambda(n)$ at level $n$:
\be
  \lambda_\infty
= \frac{\lambda(n)}{1 - \frac{2\pi C}
       {y\Gamma(h)^2}\frac{\lambda(n)}{(2\pi n)^y}
    }
+ O(\lambda^3)
\;,\;\;\;\;
  \lambda(n) 
= \frac{\lambda_\infty}{1 + \frac{2\pi C}
       {y\Gamma(h)^2}\frac{\lambda_\infty}{(2\pi n)^y}
    }
+ O(\lambda^3)
\;.
\label{eq:ccpred}
\ee
This is our one-loop prediction for the coupling constant renormalisation.
As we see below, this can be improved to take into account the level $m$
of the unperturbed state which leads to the replacement of $n$ by
$n-m$ in \eref{eq:ccpred}.

\sect{The ground-state divergence and the energy rescaling}
\label{sec:er}

\subsection{Perturbation theory results}

As in \cite{W2011}, the energy rescaling arises as the sub-leading
correction to the coupling to the identity operator. The bulk case is
not as clear-cut as the boundary case, however, as the presence of
multiple internal 
channels means that there are small differences in the rescaling for
states that arise from different representations. We shall also see
that the rescaling does not necessarily go to zero for all
renormalisable perturbations, and for the perturbation by a single
field, it diverges with $n$ if $h>3/4$.
We find these corrections by evaluating the eigenvalues of the
perturbed Hamiltonian to second order.

We consider the simplest case of the perturbation by a single field
$\varphi$ of weight $h$ with coupling $\lambda$ and the scaling
operator 
\be
  \hat h = (L_0 + \bar L_0 - \tfrac{c}{12}) 
+ \sum_i \tilde\lambda_i
   \int_{\theta=0}^{2\pi}\varphi_i(e^{i\theta})\,\frac{\rd\theta}{2\pi}
\;,
\ee
where, again, $\tilde\lambda=(2\pi)^{-y}\lambda$.
The eigenvalues of $\hat h$ are the scaling functions and 
we denote the $i$-th eigenvalue by $e_i$ and take its
expansion to be 
\be  
  e_i(\lambda)
= \sum e_{i,m}\; 
  \tilde\lambda^m
\;,
\label{eq:er0}
\ee
If $h\geq 1/2$ then one or more of these coefficients will formally be
divergent. 
For example, if the unperturbed state $\vec i$ is a highest weight
state then the first three coefficients are
\bea
&&e_{i,0} = (2 h_i - \tfrac{c}{12})
\;,\;\;
e_{i,1} = 2\pi C_{i\varphi i}
\;,\;\;
\label{eq:er1a}
\\
&&e_{i,2} = 
  - 2\pi \int_{|z|<1}\frac{\rd^2 z}{|z|^{y}}
   \left(
    \cev i\varphi(1)\varphi(z)\vec i
  - \frac{(C_{i\varphi i})^2}{|z|^{2h}}
   \right)
\;.
\label{eq:er1}
\eea
For all states except the the vacuum, $\vec i=\vac$, the integral in
$e_{i,2}$ depends in detail on the model in question but is divergent
if $h\geq 1/2$. For the vacuum case with $h<1/2$, the third and fourth
coefficients are given in \cite{KMB370} as\footnote{Note that these
  differ by powers of $(2\pi)^y$ as \cite{KMB370} uses the expansion
  parameter $\lambda$, not $\tilde\lambda$.}
\be
  e^A_{0,2} = -\tfrac 14(2\pi)^{2}\gamma^2(1 - \tfrac y2)\gamma(y-1)
\;,\;\;
  e^A_{0,3} = \frac{(2\pi)^{3}}{48}\gamma^3(\tfrac 12 - \tfrac
  14)\gamma(\tfrac{3y}4 - \tfrac 12)C_{\varphi\varphi\varphi}
\;,
\label{eq:e034}
\ee
where $\gamma(x) = \Gamma(x)/\Gamma(1-x)$ and $A$ denotes the analytic
expression. 
The expressions \eref{eq:e034} can be analytically continued to
$h>1/2$ and then agree
with the coefficients in the corresponding TBA calculation.
As pointed out in \cite{KMB370}, the TCSA method does not reproduce
the analytically continued expressions but instead approximates the
divergent expression \eref{eq:er1}. We now demonstrate how we calculate
these terms. 

\subsection{TCSA results}

The first truncation effects arise in the coefficients $e_{i,2}$. We
shall denote the contribution to $e_{i,m}$ from the states at TCSA
truncation level 
$n$ by $e_{i,m}^{[n]}$ and the full
coefficient in the exact TCSA expansion at truncation
level $n$ by $e_{i,m}^{n}$, so that the TCSA approximation to
$e_i(\lambda)$ is 
\be
e_i^n(\lambda) = \sum_m e_{i,m}^n \,\tilde \lambda^m
\;,\;\;\;\;
e_{i,m}^{n} = \sum_{k=0}^n e_{i,m}^{[k]}
\;.
\ee
The term $e_{i,2}^{[n]}$ comes  from level $n$ intermediate
states in the four-point function $\cev i \varphi(1)\varphi(z)\vec
i$. In the boundary situation in \cite{W2011} we could arrange the
boundary conditions
so that the four point function in that calculation was given by a single
chiral block; in the bulk this is not possible. Instead, 
we expand the four point function in \eref{eq:er1} over the set of
chiral blocks\footnote{See \cite{YBk} for details} as
\be
  \cev i\varphi(1)\varphi(z)\vec i
= \sum_j (C_{i\varphi j})^2 
  \left| { \cbb{2300}{i}{\varphi}{j}{\varphi}{i}{1}{z}} \right|^2
\;.
\label{eq:er2}
\ee
The contribution from the states at level $n$ in the $j$ intermediate
channel comes from the coefficient of $z^{n-h_{{i}}-h+h_j}$. As in
\cite{W2011}, we find
the leading $n$-dependence of this coefficient by expanding the
conformal block in powers of $(1-z)$,
\bea
 \cbb{2300}{{i}}{\varphi}{j}{\varphi}{{i}}{1}{z}
&=& \sum_k
  F_{jk}\cdot {\cbI{2300}{{i}}{\varphi(1)}{\varphi(z)}{{i}}{k}{7}{8}}
\\[1mm]
&=& 
  F^i_{j1}\cdot {\cbI{2300}{{i}}{\varphi(1)}{\varphi(z)}{{i}}{1}{}{}}
+ F^i_{j\varphi}\cdot {\cbI{2300}{{i}}{\varphi(1)}{\varphi(z)}{{i}}{\varphi}{}{}}
+ \ldots
\;,
%\nonumber\\
%&=& 
%  F_{j1}(1-z)^{-2h} + F_{j\varphi}(1-z)^{-h} + \ldots
%\;,
\label{eq:z1}
\eea
where $F^i_{jk}=F_{jk}[{{\varphi\varphi}\atop{ii}}]$ are the crossing
matrix elements and
\bea
 {\cbI{2300}{{i}}{\varphi(1)}{\varphi(z)}{{i}}{1}{}{}}
&=& (1-z)^{-2h}\left( 1 + \tfrac{2 h^2}{c}(1-z)^2  + O(1-z)^3 \right)
\;,\;\;
\nonumber\\[2mm]
 {\cbI{2300}{{i}}{\varphi(1)}{\varphi(z)}{{i}}{\varphi}{}{}}
&=& (1-z)^{- h}\left(1 + \tfrac{h}{2} (1-z) + O(1-z)^2 \right)
\;.
\eea
We have assumed that the identity and $\varphi$ are the most singular
fields in the operator product 
$\varphi*\varphi$; if not, there will be correspondingly more terms in
\eref{eq:z1}. 
Hence we find the leading terms arising at truncation level $n$ in the
chiral block are
\bea
&&
%\left. 
{\cbb{2300}{{i}}{\varphi}{j|_n}{\varphi}{{i}}{1}{z}}
%\right|_n
\nonumber\\[3mm]
&=& 
  z^{n-h_{i}-h+h_j}
  \left(
  \frac{F^i_{j1}\Gamma(n{-}h_{i}{+}h_j{+}h)}
             {\Gamma(2h)\Gamma(n{-}h_{i}{+}h_j{-}h{+}1)}
+
  \frac{F^i_{j\varphi}\Gamma(n{-}h_{i}{+}h_j)}
             {\Gamma(2h)\Gamma(n{-}h_{i}{+}h_j{-}h{+}1)}
+\ldots
  \right)
\eea
%\newpage
This gives the second order truncation level $n$ contribution to the energy
as
\bea
%&&
e^{[n]}_{i,2}
%\nonumber\\
&{=}& - 2\pi
    \sum_j 
\frac{(C_{{i}\varphi j})^2}
     {2(n{-}h_{i}{+}h_j)\Gamma(n{-}h_{i}{-}h{+}h_j{+}1)^2}
\nonumber\\
&&\qquad\qquad\qquad\qquad{}\times \left[
  F^i_{j1} \frac{\Gamma(n{-}h_{i}{+}h{+}h_j)}{\Gamma(2h)}
+ F^i_{j\varphi} \frac{\Gamma(n{-}h_{i}{+}h_j)}{\Gamma(h)}
+ \ldots
\right]^2
\nonumber\\
&{=}& -2\pi
    \sum_j (C_{{i}\varphi j})^2 \Big[
    n^{4h{-}3}\frac{(F^i_{j1})^2}{2 \Gamma(2h)^2}
 +  n^{4h{-}4}(4h-3)\frac{(F^i_{j1})^2 (h_j{-}h_{i})}{2 \Gamma(2h)^2}
\nonumber\\
&&\qquad\qquad{}
 +  n^{3h{-}3}\frac{F^i_{j1}F^i_{j\varphi}}{\Gamma(h)\Gamma(2h)}
 +  n^{2h{-}3}\frac{(F^i_{j\varphi})^2}{2\Gamma(h)}
 +  O(n^{4h{-}5},n^{3h{-}4},n^{2h{-}4})
    \Big]
\;.
\label{eq:er4}
\eea
Using the crossing properties of the full correlation functions, we
find
\be
  \sum_j (F^i_{j1} C_{{i}\varphi j})^2 = 1
\;,\;\;
  \sum_j (F^i_{j\varphi} C_{{i}\varphi j})^2 
= C_{\varphi\varphi\varphi} C_{{i}{i}\varphi}
\;,\;\;
  \sum_j F^i_{j1} F^i_{j\varphi} (C_{{i}\varphi j})^2 
= 0
\;,
\ee
so that the second order truncation level $n$ contributions \eref{eq:er4} are
\[
e_{i,2}^{[n]}=
 -2\pi
    \Big[
    \frac{n^{4h{-}3}}{2 \Gamma(2h)^2}
 +  (4h-3)\frac{(\alpha_i{-} h_{i})n^{4h{-}4}}
                     {2 \Gamma(2h)^2}
%\nonumber\\
%&&\qquad\qquad\qquad {} 
 +  \frac{C_{\varphi\varphi\varphi} C_{{i}{i}\varphi}}
                   {2\Gamma(h)}n^{2h{-}3}
 +  O(n^{4h{-}5},n^{3h{-}4},n^{2h{-}4})
    \;\Big]
\;,
\]
\be
%&&
e_{0,2}^{[n]}=
 -2\pi
    \Big[
    \frac{n^{4h{-}3}}{2 \Gamma(2h)^2}
 +  (4h-3)\frac{h n^{4h{-}4}}
                     {2 \Gamma(2h)^2}
 +  O(n^{4h{-}5},n^{3h{-}4},n^{2h{-}4})
    \;\Big]
\;,
\ee
where 
\be
\alpha_i = \sum_jh_j( C_{{i}\varphi j}F^i_{j1})^2
= \frac{S_{11}}{S_{1\varphi}}\sum_{j\in i*\varphi} 
  \frac{S_{1j}}{S_{1i}}h_j
\;,\;\;
\alpha_0 = h
\;,
\ee
where the second expression in the modular S-matrix $S_{ij}$ holds for
a diagonal modular invariant \cite{Runkel}.
If $\vec i$ is a highest weight state then we can then sum these
contributions to get 
\bea
 e_{0,2}^n
&=&
 e_{0,2}^A - 2\pi\left[ 
   \frac{n^{4h-2}}{4(2h-1)\Gamma(2h)^2}
+ \frac{n^{4h-3}}{\Gamma(2h)^2}\frac{2h{+}1}{4}
+ \ldots
  \right]
\;,
\label{eq:er5}
\\
 e_{i,2}^n
&=&
 e_{i,2}^A - 2\pi\left[
  \frac{n^{4h-2}}{4(2h-1)\Gamma(2h)^2}
+ \frac{n^{4h-3}}{\Gamma(2h)^2}\frac{2(\alpha_i{-}h_i){+}1}{4}
+ \frac{C_{ii\varphi}C_{\varphi\varphi\varphi}n^{2h-2}}
  {2(2h-2)\Gamma(h)^2}
+\ldots
 \right]
\;.
\nonumber\eea
If the TCSA coefficients are convergent, then \eref{eq:er5} gives the
leading corrections; if they are divergent, they give the leading
divergences. 
We first consider this result for the ground state scaling function.

For $h>1/2$, the leading term(s) in \eref{eq:er5} are divergent. To
get a proper estimate of the ground state energy from TCSA we need
explicitly to subtract these. We do this for the divergent case
considered in \cite{LCM1} in section \ref{sec:lcm}. As will be seen,
this is very effective at producing a good estimate from the TCSA.

For excited states, we shall only consider the scaled energy gap,
$\tilde e_i = e_i - e_0 = \sum \tilde e_{i,k}\tilde\lambda^k$ and 
its TCSA approximation
$\tilde e^n_i = \sum \tilde e^n_{i,k}\tilde\lambda^k$.
Using \eref{eq:er5}, we can write the TCSA approximation as
\bea
  \tilde e_i^n(\lambda)
&=& (2h_i) 
+ (2\pi)\tilde\lambda C_{i\varphi i}
+ \tilde\lambda^2\left(
  \tilde e^A_{i,2} 
   + 2\pi\frac{n^{1-2y}}{\Gamma(2h)^2}
                          \frac{h_i - \delta_i}{2} 
   + 2\pi C_{ii\varphi}C_{\varphi\varphi\varphi}
     \frac{n^{-y}}{2y\Gamma(h)^2}
  \right)
\nonumber
\\&=&
 2\delta_i {+} 
 \left(1 {+} 
           \frac{2\pi\tilde\lambda^2}{4\Gamma(2h)^2n^{2y-1}}
 \right)
 \left( 2(h_i{-}\delta_i) 
     {+} 
    (2\pi) \left( \tilde\lambda + 
     \frac{\tilde\lambda^2 C_{\varphi\varphi\varphi}}{2y\Gamma(h)^2n^y}\right)
                C_{i\varphi i}
             + \tilde e^A_{0,2}\tilde\lambda^2
\right) {+}\ldots 
\nonumber
\\&=&
  r_n^i(\lambda)\; \left[\tilde e_i( \lambda g_n(\lambda))
    -2\delta_i\right] + 2\delta_i
  \;+\; O(\lambda^3)
\;,
\eea
where $\delta_i = \alpha_i - h_i$.
which defines the one-loop energy rescaling $r_n^i$ and coupling constant
renormalisation $g_n(\lambda)$,
\be
 r^i_n(\lambda) = \left(1 + 
           \frac{\lambda^2}{4\Gamma(2h)^2(2\pi n)^{2y-1}}
 \right)
\;,\;\;
 g_n(\lambda) = 
\left( 1 + 
     \frac{\lambda \,C}{2y\Gamma(h)^2(2\pi n)^y}\right)
\;.
\label{eq:rngndef}
\ee
The coupling constant renormalisation agrees perfectly with the
result of the previous section, \eref{eq:ccpred}. The energy rescaling
is a new prediction. It differs from the boundary case as the energy
rescaling is not exactly the same for each state, because there is at
this order a small shift $\delta_i$ which differs between states
arising in different representations.
Since this is a small overall constant shift in each representation,
it is only important for the lowest lying states in each
representation where it does make a noticeable difference, as we see
in the plots in section \ref{sec:km}.

These results also hold, suitably adjusted, for excited states. If
$\vec\psi$ is a state at level $m$ in
the representation $i$ then the contribution in 
\eref{eq:er2} from the intermediate states at level $n$ comes from the
coefficient of $z^{n-(h_{{i}}+m)-h+h_j}$ and following the changes
through the net result is to replace $n$ in \eref{eq:rngndef} by
$(n-m)$, so that for the state $\vec\psi$ the rescaling and
renormalisation are 
\be
 r^\psi_n(\lambda) = \left(1 + 
           \frac{\lambda^2}{4\Gamma(2h)^2(2\pi (n-m))^{2y-1}}
 \right)
\;,\;\;
 g_n^\psi(\lambda) = 
\left( 1 + 
     \frac{\lambda \,C}{2y\Gamma(h)^2(2\pi (n-m))^y}\right)
\;.
\label{eq:rngndef2}
\ee
This alteration is a sub-leading effect in $n$ but is appreciable for
the cases we consider when $n$ is sometimes quite small.

%\newpage
\sect{Tests of the TCSA in the tri-critical Ising model}
\label{sec:tcim}

One of the first paper to use the TCSA extensively was
\cite{LCM1}. This investigated perturbations of the tri-critical Ising
model with mixed success. With hindsight, it is easy to see now why
they obtained good results for the perturbations by the fields of
weight $1/10$ and $3/80$ (with fast convergence), mixed results for
the field of weight $7/16$ (with slow convergence) and very poor
results for the massless perturbation by the field of weight $3/5$.
We reconsider this final case in the light of our results in section
\ref{sec:lcm}. 
Klassen and Melzer also used the TCSA to test predictions of the
IR scattering description of the massive perturbation by the field of
weight $3/5$ in \cite{KM1}. They too found rather poor agreement with
TCSA -- we are able to improve on it greatly using the renormalisation
and re-scalings, which we show in section \ref{sec:km}.

Following \cite{KMB370}, we denote the massless perturbation of the
tri-critical Ising model by $\Mc A_4^{(+)}$ and the massive perturbation
by $\Mc A_4^{(-)}$. 
These have been investigated extensively and TBA systems found for the
ground state(s) in \cite{Zamo9} and \cite{Zamo9b} respectively.
In both cases once can define a mass-scale $m$ and a dimensionless
system size $r = mR$ satisfying
\be
  r = \frac{56 (21 \pi)^{1/4}  }{25 \sqrt{5}}
     \left(\frac{\Gamma
   \left(-\frac{7}{5}\right) \Gamma \left(\frac{1}{5}\right)}{\Gamma
   \left(\frac{4}{5}\right) \Gamma
   \left(\frac{12}{5}\right)}\right)^{5/8}
  |\lambda |^{5/4}
 = 10.83\ldots\, |\lambda|^{5/4}
\;.
\label{eq:rdef}
\ee
For $\Mc A_4^{(-)}$, $m$ is in the mass of the kink in this model; for
$\Mc A_4^{(+)}$ it is an inherent mass-scale in the problem and can
be identified with the kink mass by analytic continuation in $\lambda$.

\subsection{The ground state energy in $\Mc A_4^{(+)}$}
\label{sec:lcm}

\begin{figure}[thb]
\subfigure[The bare data]{\scalebox{0.43}{\includegraphics{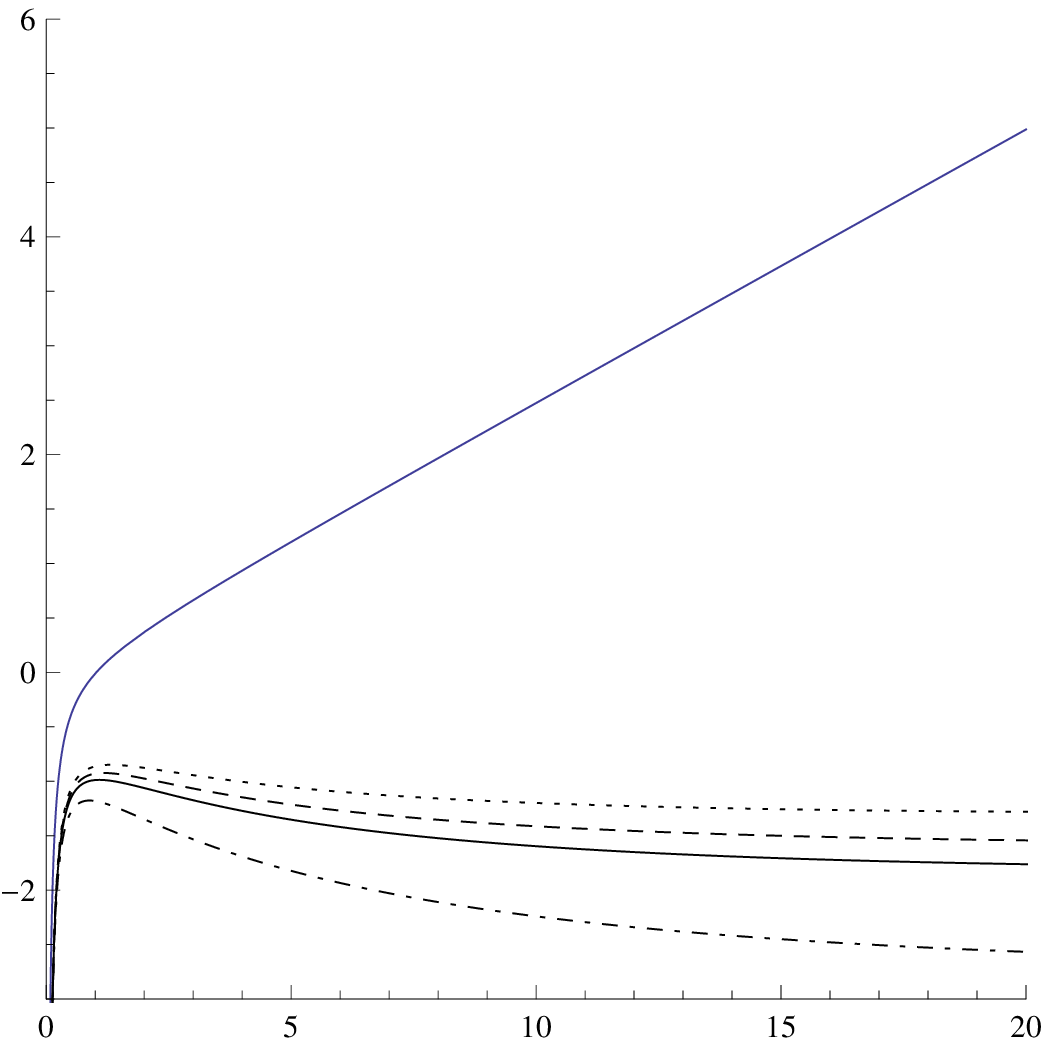}}
\label{fig:cp1}}
\hfill
\subfigure[The subtracted data]{\scalebox{0.43}{\includegraphics{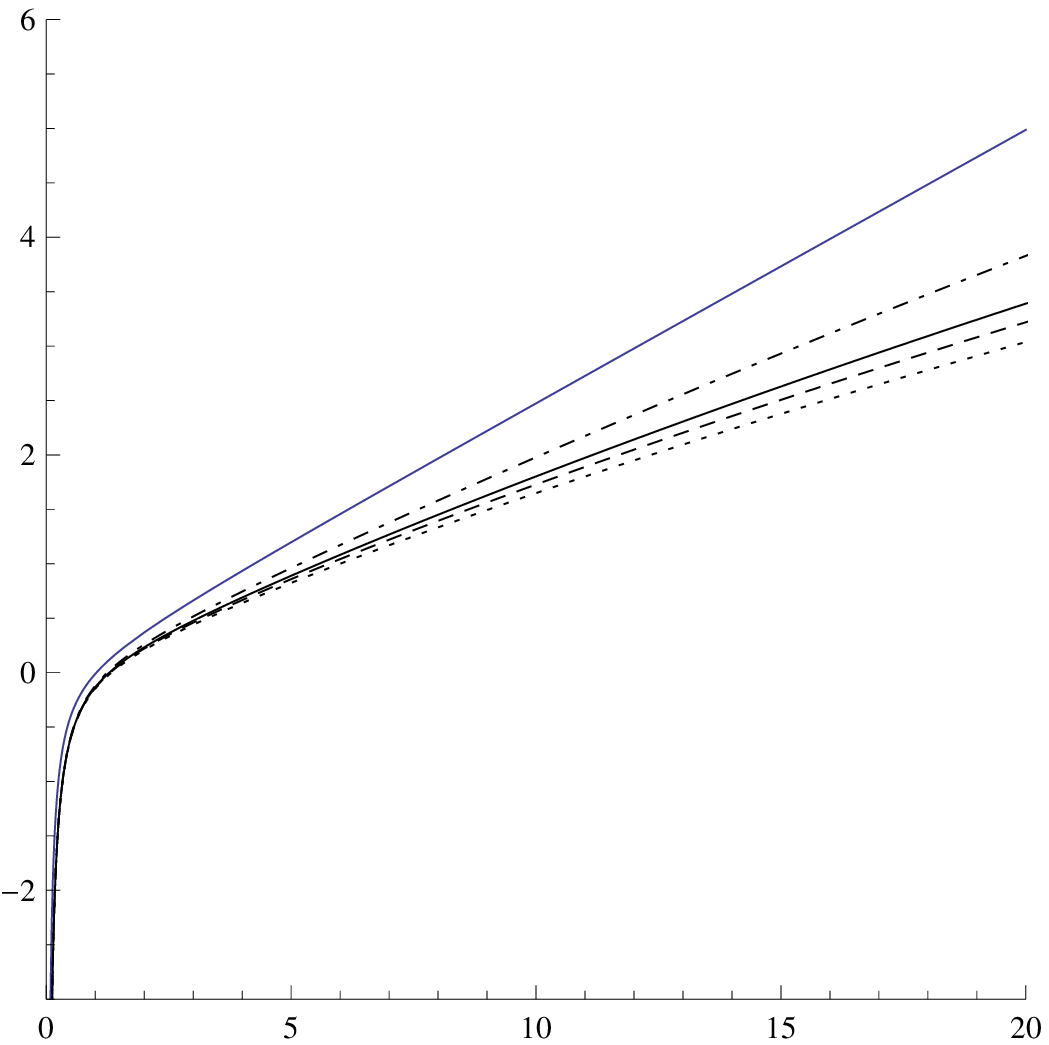}}
\label{fig:cp2}}
\hfill
\subfigure[The subtracted, renormalised data]{\scalebox{0.43}{\includegraphics{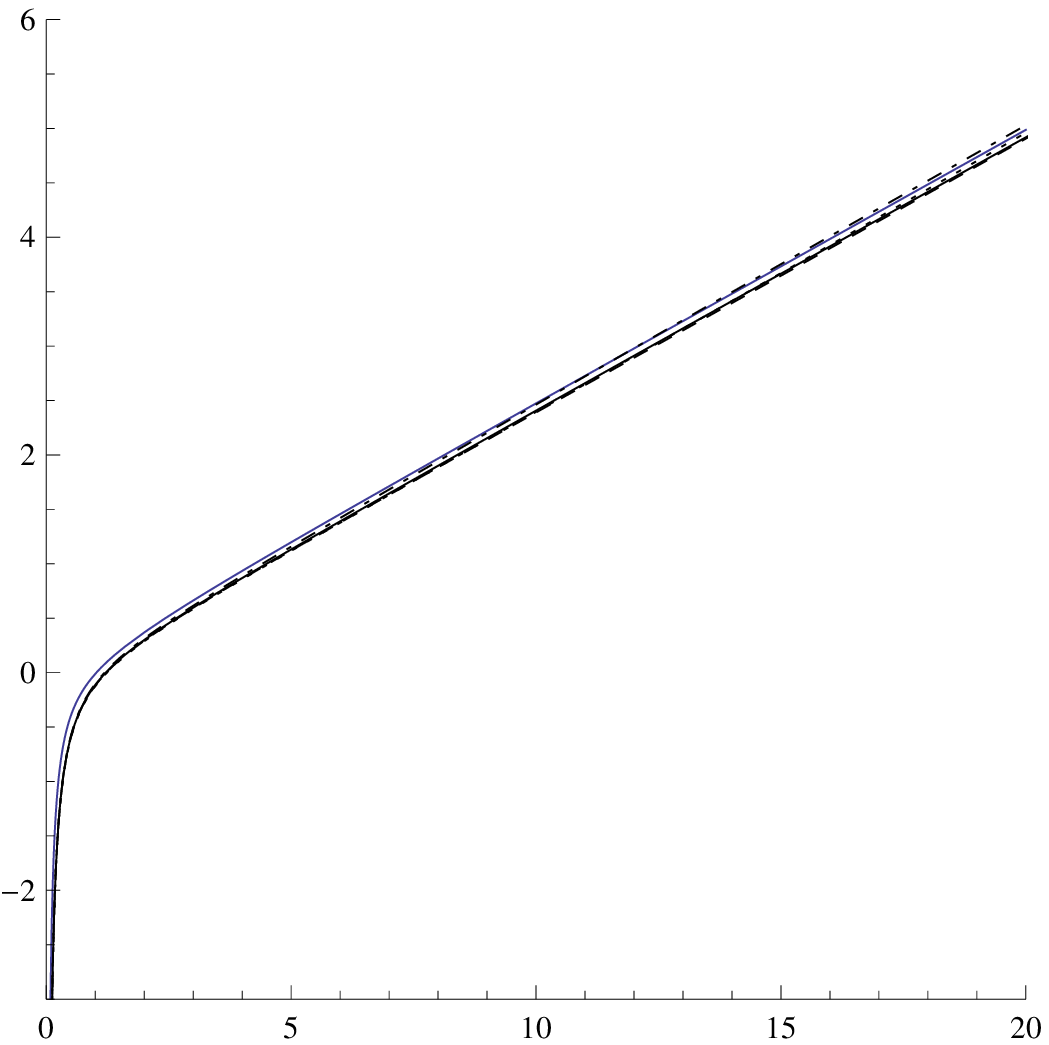}}
\label{fig:cp3}}
\caption{The ground state energy for the model $\Mc A_4^{(+)}$ at
truncation levels 3 (dotted), 4 (dashed), 5 (solid) and 9 (dot-dashed)
plotted against $r$ together with the leading exact IR
behaviour (thin solid line). 
} 
\label{fig:cp123}
\end{figure}

In figure \ref{fig:cp1} we show the TCSA estimates of the ground state
energy $(2\pi/r)e_0(r)$ for the massless perturbation $\Mc A_4^{(+)}$ plotted
against $r$ for truncation levels 3, 4 and 5. 
As L\"assig et al observed, the ground state energy does not appear to
be converging to the expected linear behaviour of a massless theory.
This was explained in \cite{KMB370} as a consequence of the divergence
of the ground state energy.

In figure  \ref{fig:cp2} we show the same TCSA data but with the
leading divergence and sub-leading correction given in \eref{eq:er5}
subtracted, that is, we plot
\be
  \left(\frac{2\pi}{r}\right)
  ( e^n_0(\lambda)  
 - (2\pi)^{2-2y}\lambda^2\left[ 
   \frac{n^{4h-2}}{4(2h-1)\Gamma(2h)^2}
+ \frac{n^{4h-3}}{\Gamma(2h)^2}\frac{2h{+}1}{4}
  \right]
  )
\;,
\ee
against $r$, where $h=3/5$ and $r$ is given in \eref{eq:rdef}.
We also plot the expected IR behaviour which is 
\be
 -\frac{\pi}{12 r} + \frac{r}{4} 
\;,
\ee
where the linear term can be deduced from the perturbative expansion
of the TBA solution given in \cite{Zamo9}.

As is clearly seen in figure \ref{fig:cp123}, after subtraction of the
leading divergence, the TCSA data appears to be converging towards the
expected IR behaviour.
This convergence is dramatically improved by incorporation of the
leading coupling constant renormalisation as in figure \ref{fig:cp3}.
Since this is the ground-state, there is no energy rescaling to include.

%\newpage
\subsection{The energy gaps in $\Mc A_4^{(+)}$}

The scaling function gaps in $\Mc A_4^{(+)}$ were also investigated in
\cite{LCM1}. They found poor convergence of the first even excitation,
but good evidence of the IR fixed point being the Ising model. 
We report here that in fact we have found rather different convergence of
the first even gap to that presented in \cite{LCM1}, something
we find hard to explain.

\begin{figure}[thb]
\centering{
\subfigure[The bare TCSA data]{\scalebox{0.455}{\includegraphics{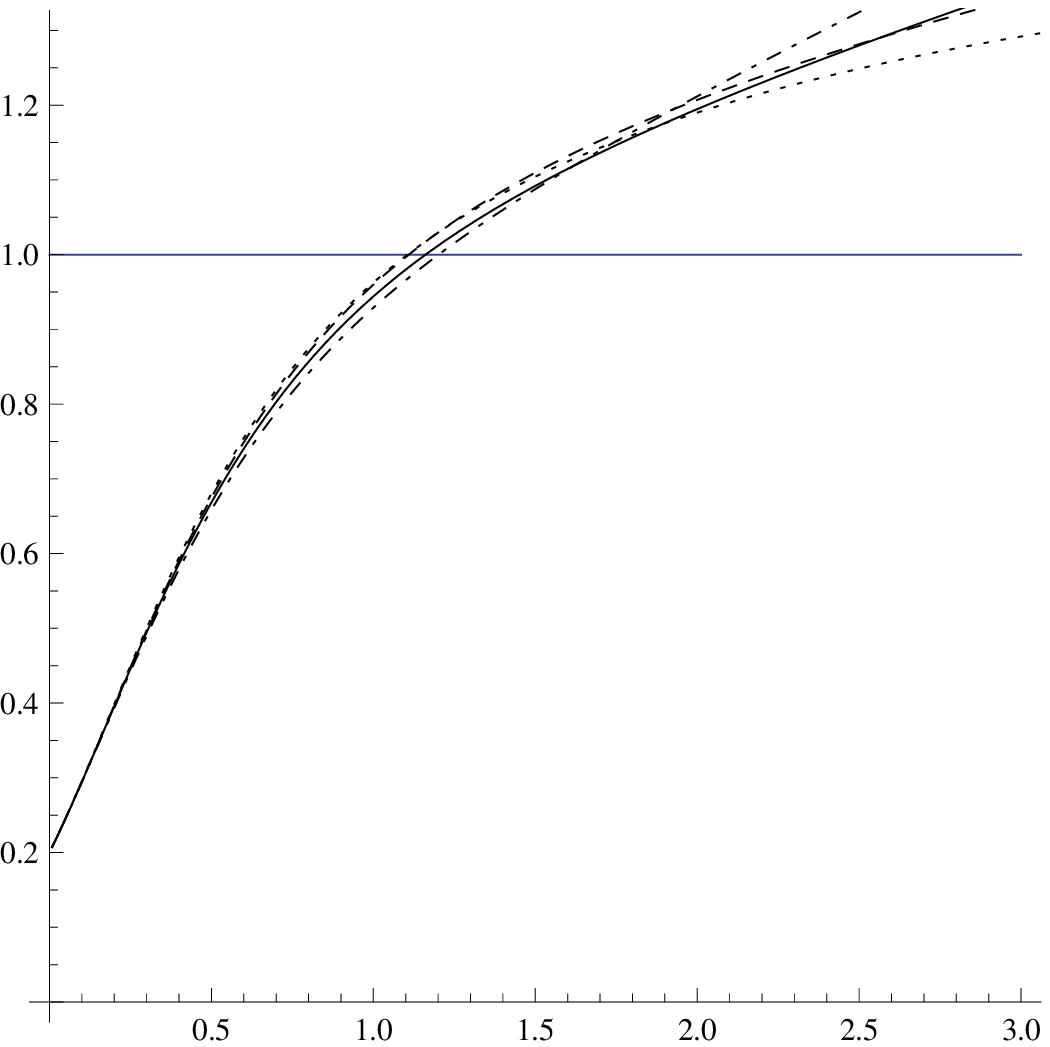}}\label{fig:cp4}
}
\hfil
\subfigure[The renormalised, rescaled TCSA data]{\scalebox{0.455}{\includegraphics{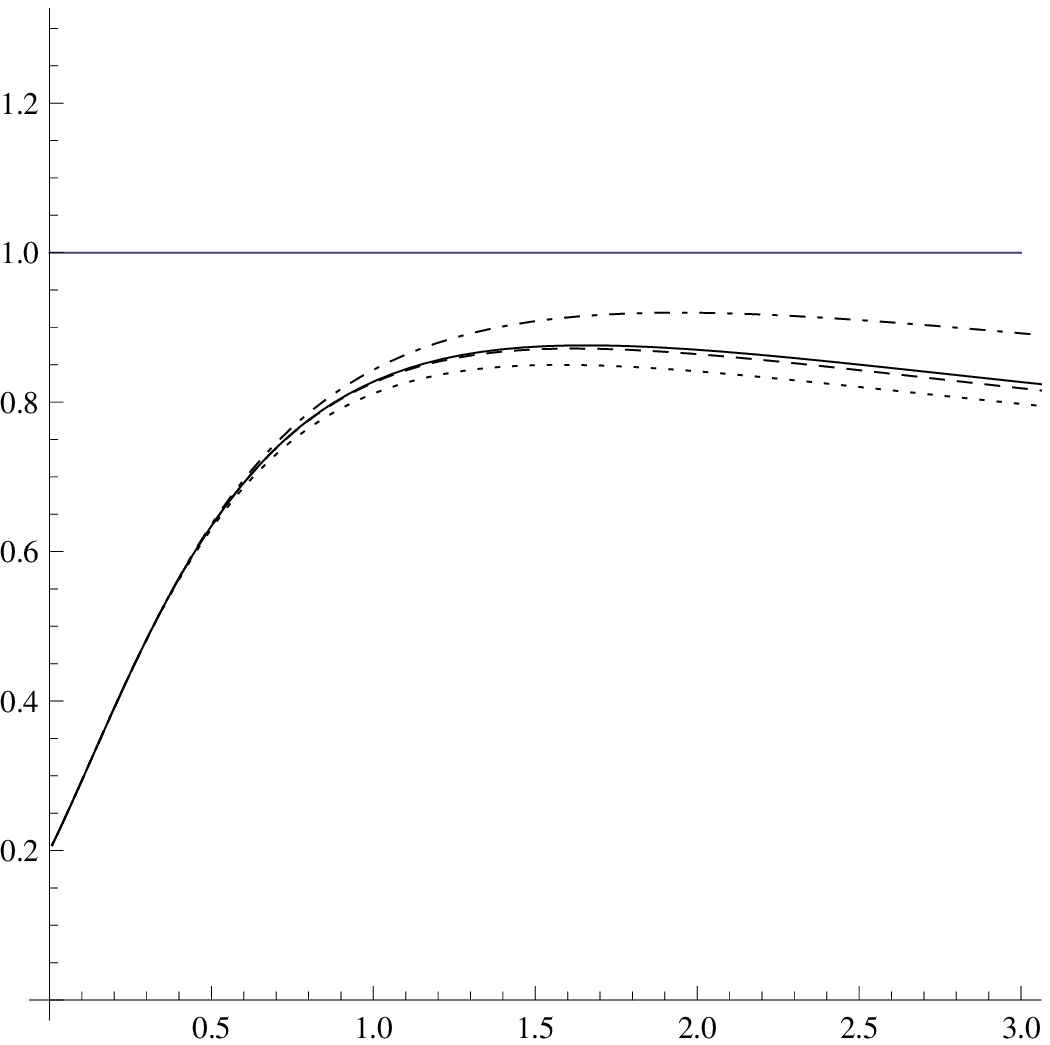}}\label{fig:cp4b}
}
}
\caption{The first gap in the even sector of $\Mc A_4^{(+)}$ at
  truncation levels 3 (dotted), 4 (dashed), 5 (solid) and 9
  (dot-dashed), plotted against $\lambda$.}
\label{fig:cp44b}
\end{figure}

In figure \ref{fig:cp4} we show the bare TCSA data, which is to be
compared with figure 13(b) of \cite{LCM1}. As can be seen, the bare TCSA data
appears to be converging slowly to its IR limit $e = 1$ for
$\lambda \lesssim 2$ but diverging for
$\lambda \gtrsim 2$; this is in accord with the idea of a fixed
point at a finite value of $\lambda$ which can be read off from
\eref{eq:rngndef} and convergence of the raw TCSA data only below this
fixed point. We also note that our TCSA data is certainly
changing more slowly with $n$ than was found in \cite{LCM1}. As
before, the 
convergence is dramatically improved by including the 1-loop rescaling
and renormalisation as shown in figure \ref{fig:cp4b}. The putative
fixed point is pushed far off to the right by the renormalisation and
the rescaling brings the data down below its $IR$ value.

\begin{figure}[thb]
\centering{
\subfigure[{The first 19 gaps in the $(1,*)$ sector}]{\scalebox{0.4}{\includegraphics{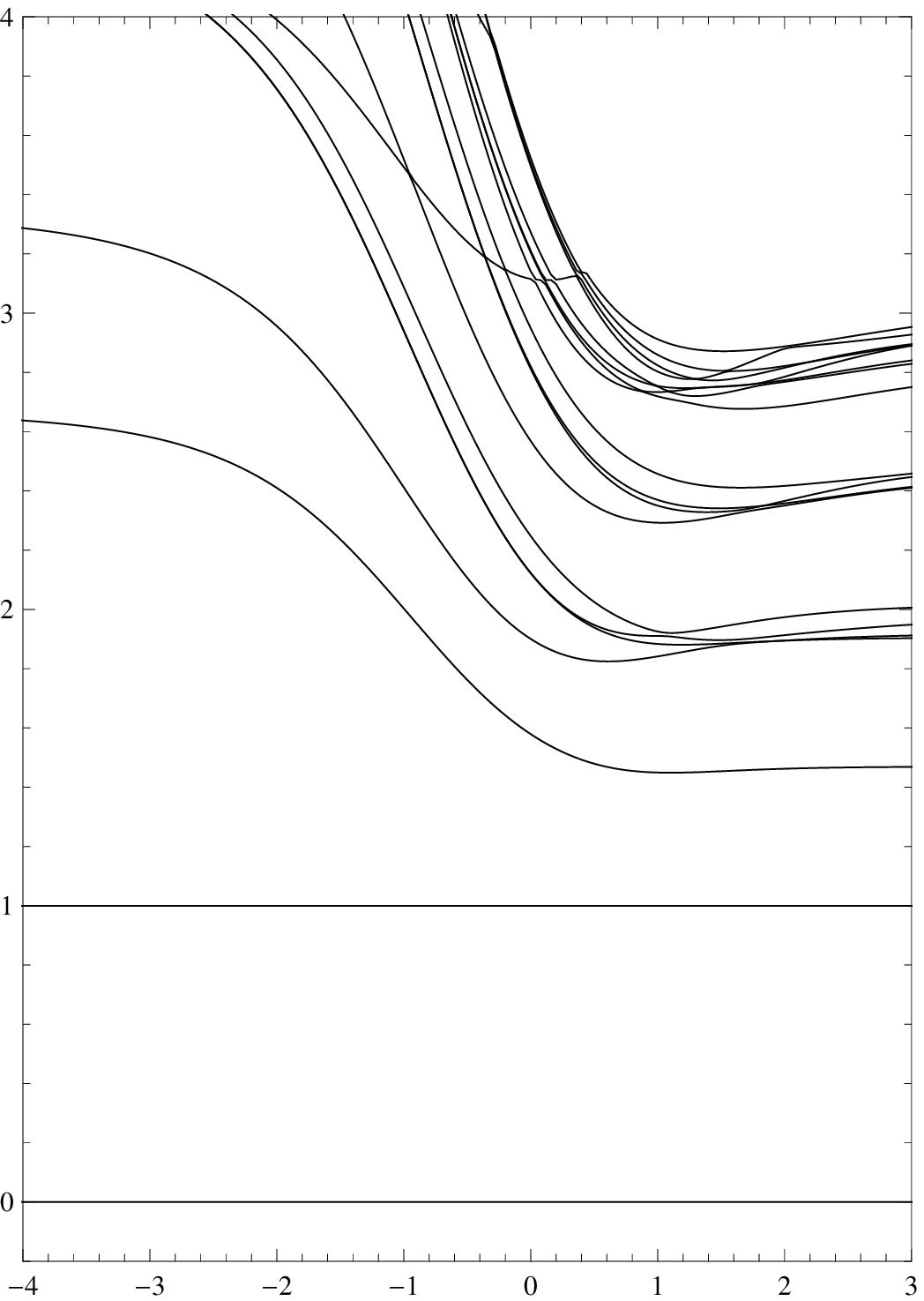}}\label{fig:cp5}}
\qquad
\subfigure[{The first 15  gaps in the $(2,*)$ sector}]{\scalebox{0.4}{\includegraphics{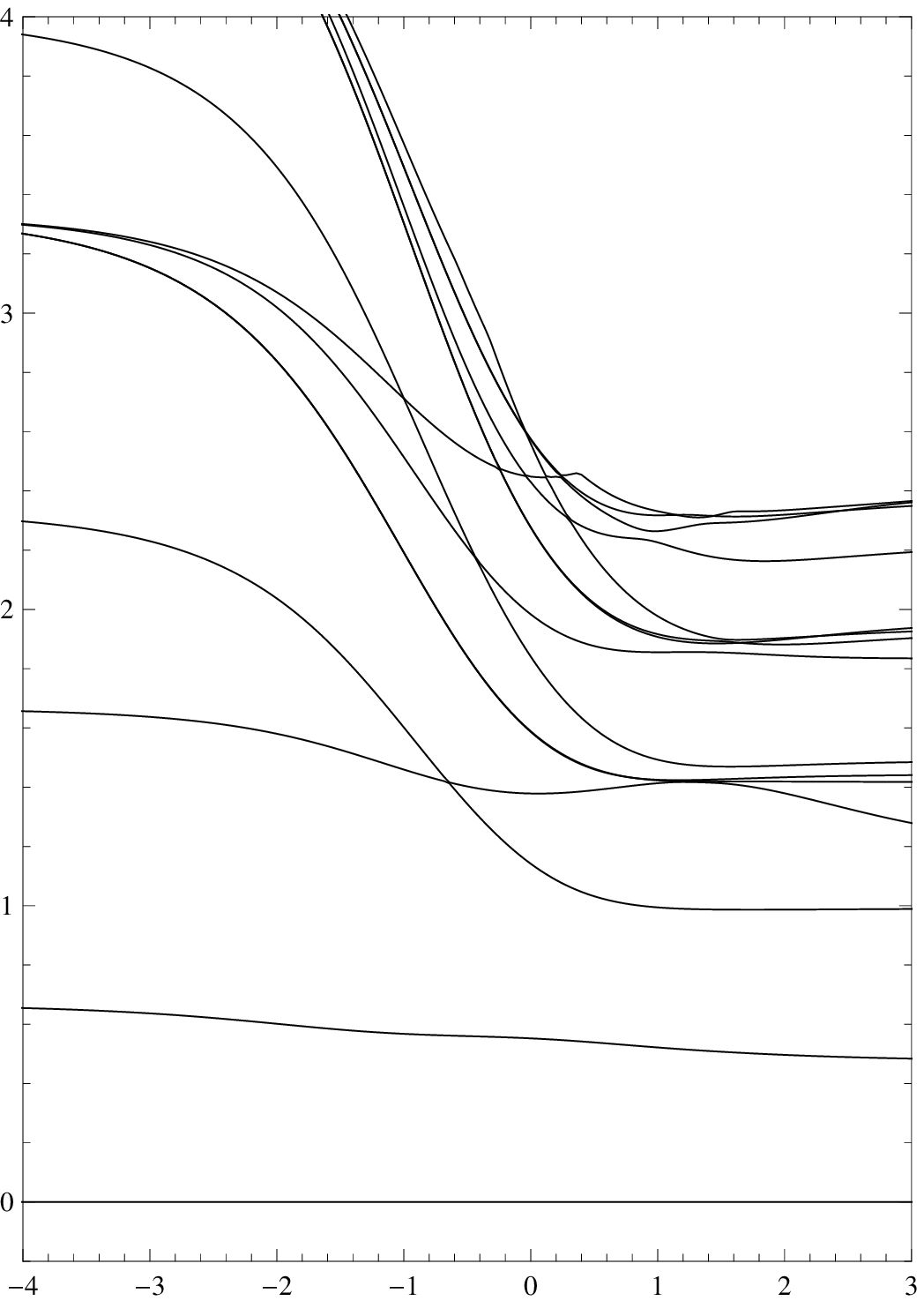}}\label{fig:cp5b}}
\qquad
\subfigure[{The first 12  gaps in the $(3,*)$ sector}]{\scalebox{0.4}{\includegraphics{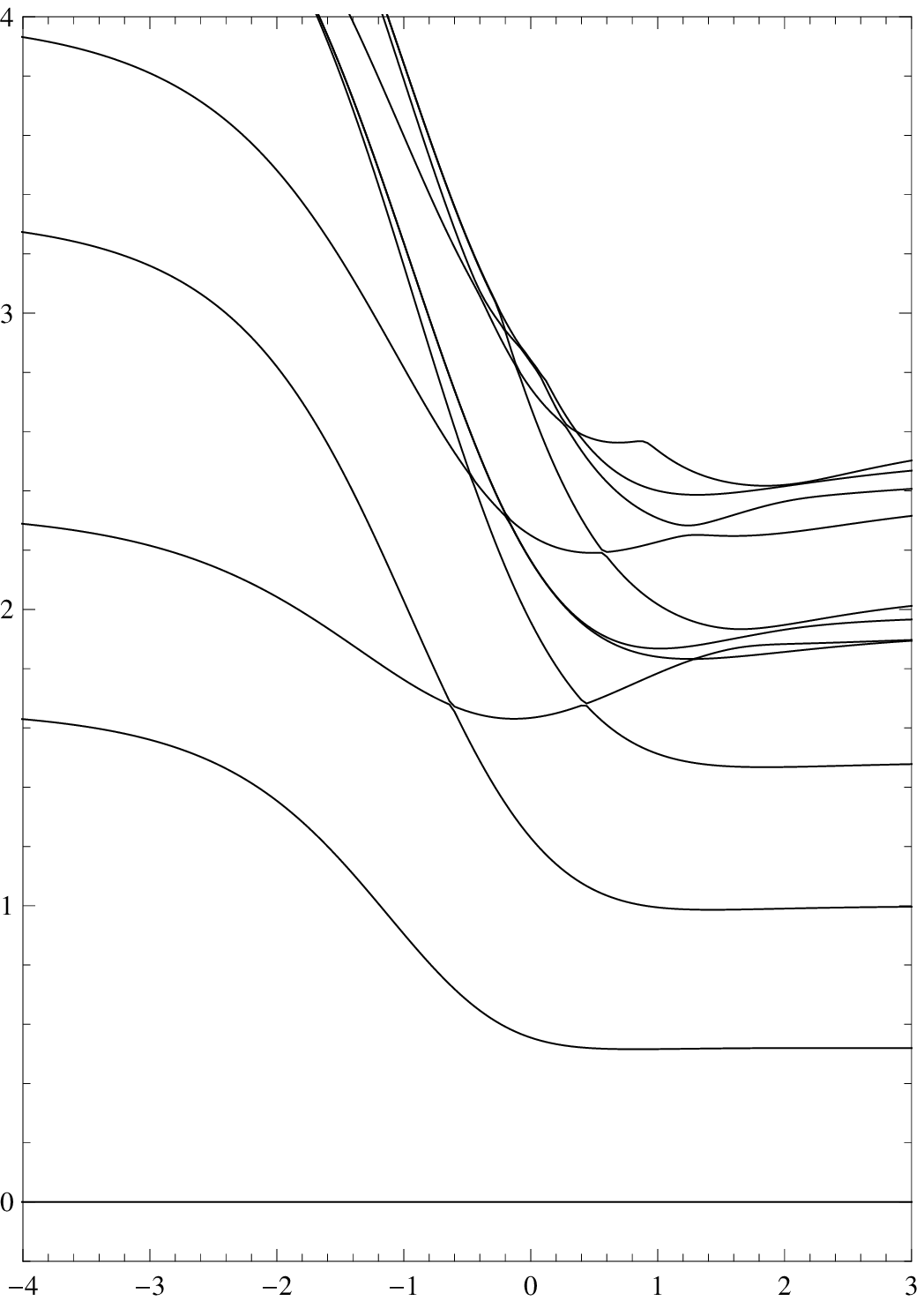}}\label{fig:cp5c}}
}
\caption{Normalised energy gaps for $\Mc A_4^{(+)}$ plotted against
  $\log(\lambda)$ at truncation level 9.}
\label{fig:cp55b}
\end{figure}

We would also like to check that the TCSA method gives good results
even when the renormalisation and re-scalings have become large. To
this end, in figure \ref{fig:cp55b}
we also include a plot of the normalised energy gaps in the $(r,*)$
sectors at truncation level 9. We have normalised these relative to
the first gap in th $(1,*)$ sector. Assuming that the IR values of the
gaps are reached around $\log(\lambda)=2$ and that the first gap in the
$(1,*)$ sector 
tends to the scaling value 4 consistent with the identification of the
state in the IR as $L_{-2}\bar L_{-2}\vac$, then the rescaling
function $r_{9}(\lambda^*)\sim 2.8$, which is definitely outside
the perturbative regime,
They also appear to show that the Ising fixed point is actually
reached at about $\log\lambda\sim 2$,  
in agreement with the prediction from the one-loop calculation of a
fixed point at a finite  positive $\lambda$. For $n=9$, the 1-loop
prediction for the position of the fixed point is $\log(\lambda)= 2.45$ .

Despite the size of the rescaling (and the presumably infinite coupling
renormalisation if the IR fixed point is actually reached)
these figures
they show good qualitative evidence for a flow to the Ising
model, the $(r,*)$ sector flowing to the $(1,r)$ representation.

%\newpage
\subsection{The energy gaps in the massive perturbation $\Mc A_4^{(-)}$.}
\label{sec:km}

The TCSA was also used in \cite{KM1} in an attempt to verify the
description of the IR limit of the massive perturbation $\Mc
A_4^{(-)}$ by a kink model and the approximation of the energy levels
using the Bethe-Yang equations (we refer the interested reader to
\cite{KM1} for details). 
The Bethe-Yang (or BY) equations give approximate finite-size energy gaps for
massive models and agree with the exact TBA results up to
corrections which are exponentially suppressed for large $r$. 

\begin{figure}[thb]
\subfigure[The bare data]{\scalebox{0.40675}{\includegraphics{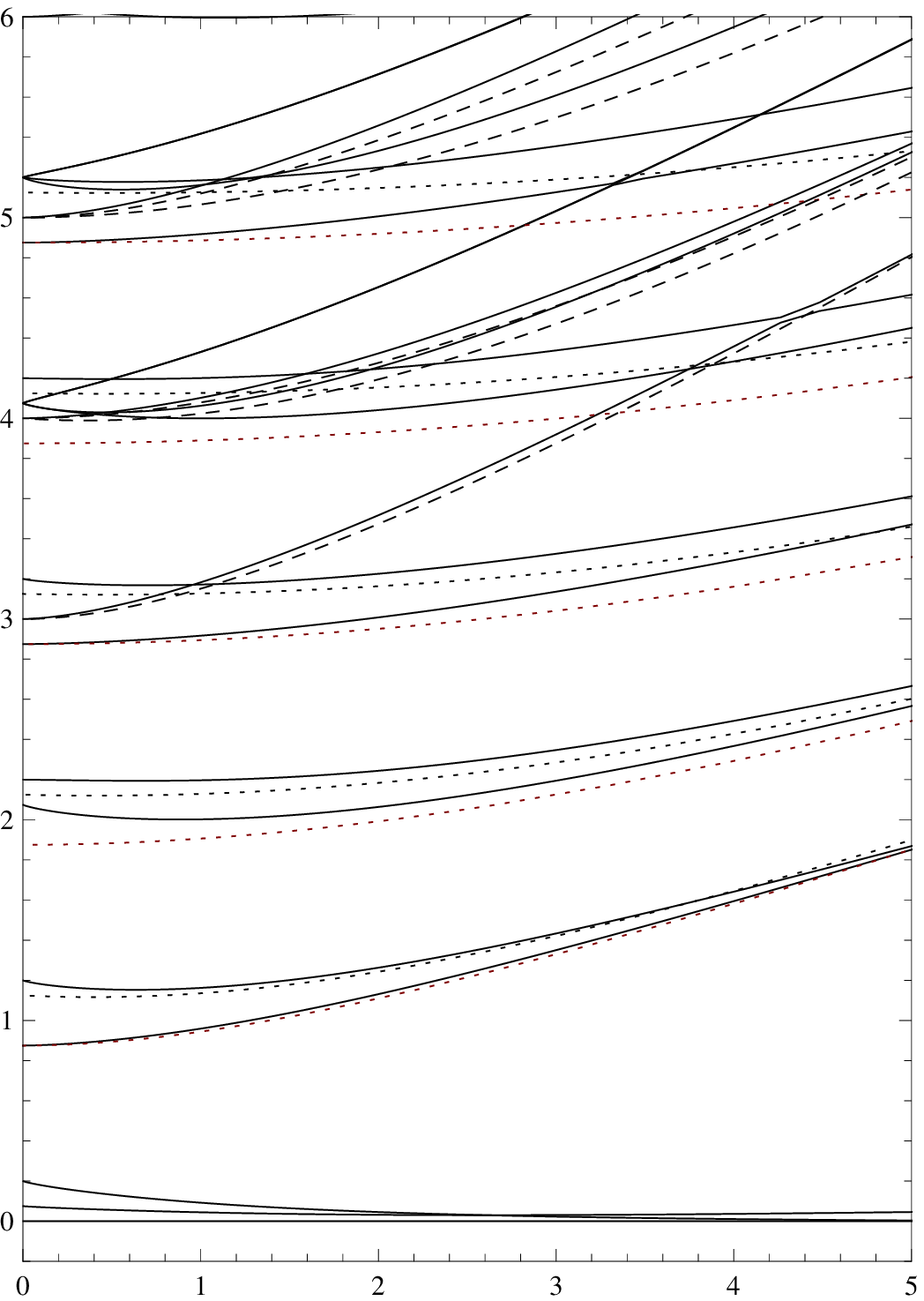}}
\label{fig:kmp1}}
\hfill
\subfigure[The renormalised, rescaled data]{\scalebox{0.40675}{\includegraphics{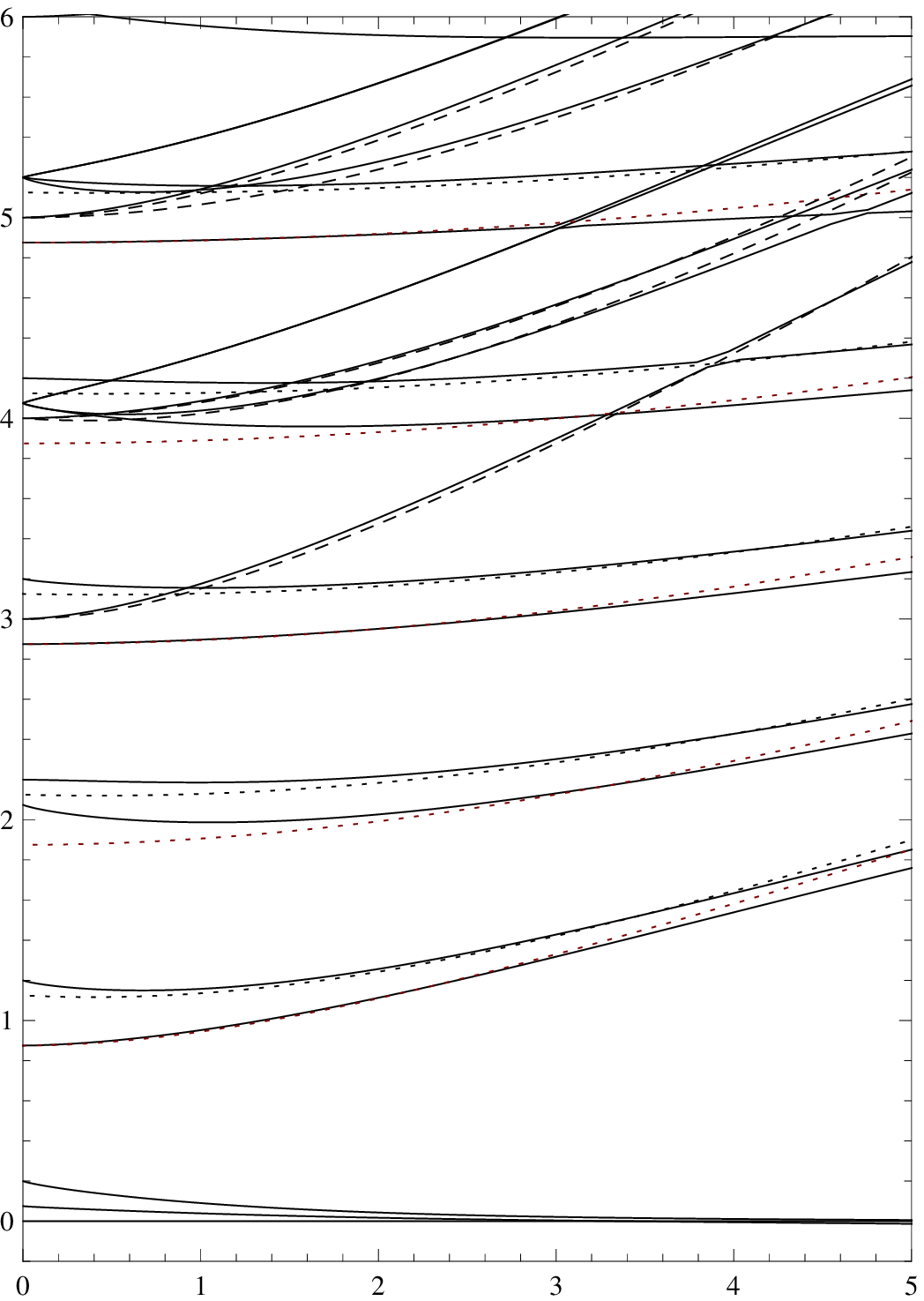}}
\label{fig:kmp2}}
\hfill
\subfigure[The fitted data]{\scalebox{0.40675}{\includegraphics{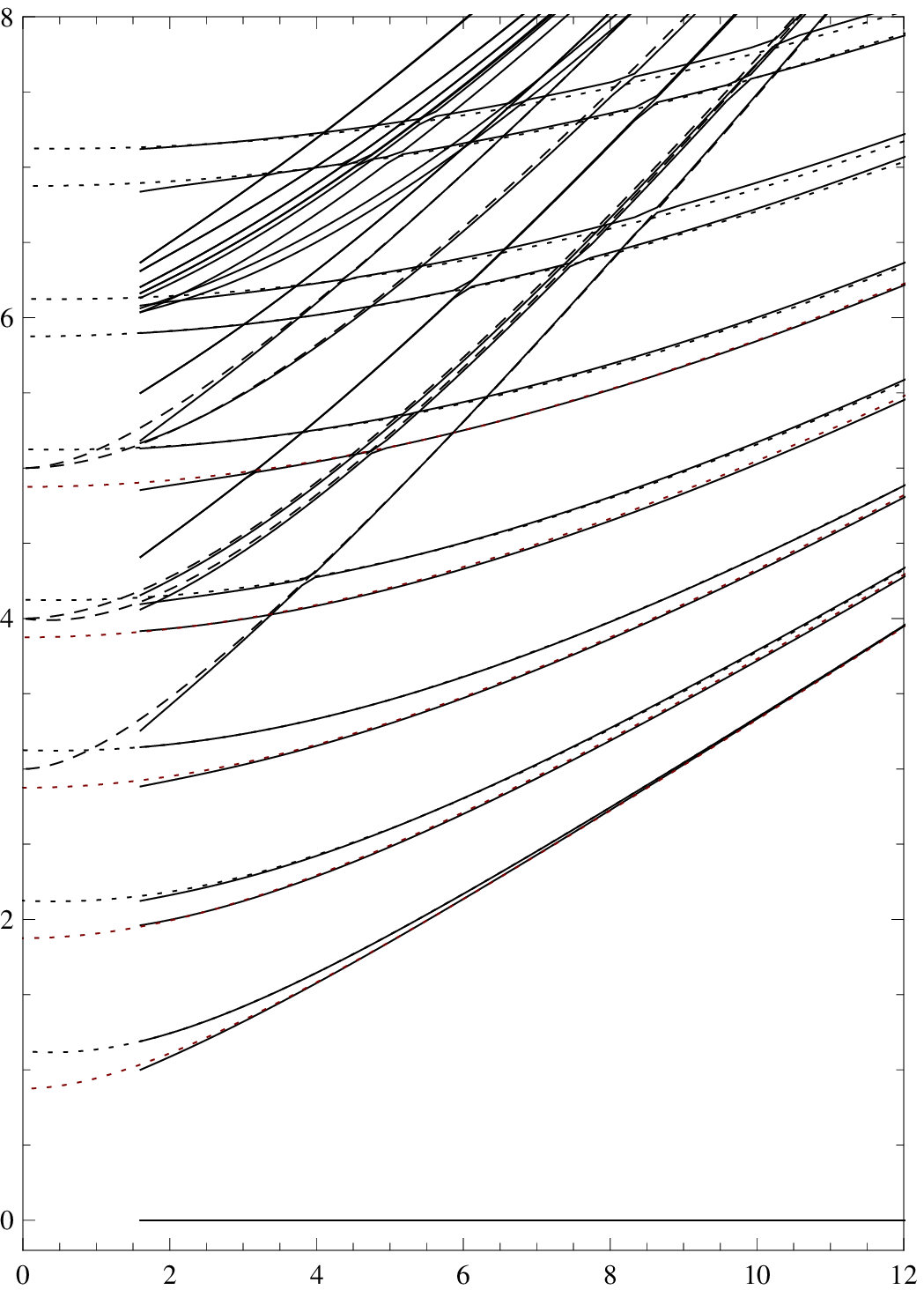}}
\label{fig:kmp3}}
\caption{The gaps for the massive perturbation $\Mc A_4^{(-)}$ at
truncation levels 5 for figures (a) and (b) and 9 for figure (c).
In all cases the
approximate Bethe-Yang two- and four-particle energies are also
given as dotted and dashed lines, as
calculated in \cite{KM1}.
} 
\label{fig:kmp123}
\end{figure}

In figure \ref{fig:kmp1}, we reproduce figure 7 from \cite{KM1} showing
the scaling functions plotted against $r$. The solid lines 
the bare TCSA data and the various dotted and dashed lines are the
BY approximations to these energy levels. There is a general
agreement, but we do not think it is good enough to confirm the
BY energy levels as correct.
As Klassen and Melzer say, the main problem is not with the BY levels
but with truncation errors in the TCSA.
In figure \ref{fig:kmp2}, we correct the TCSA data by the 1-loop
renormalisation and rescaling formulae we have found.

There is a dramatic increase in agreement with the BY results for the
renormalised, re-scaled gaps compared to the bare gaps; they
still 
disagree for small $r$ as is to be expected for the approximate BY
solutions, but the agreement for middling ranges of $r$, say $2<r<4$, is
excellent. In this range, the exponential corrections to the BY
solutions have been suppressed and the 2-loop corrections to the
renormalisation and rescaling formulae are still small. For $r>4$
the 2-loop corrections to the renormalisation and re-scalings are large
enough to show a qualitative difference between the TCSA and BY gaps.

\begin{figure}[thb]
\subfigure[The bare data]{\scalebox{0.7}{\includegraphics{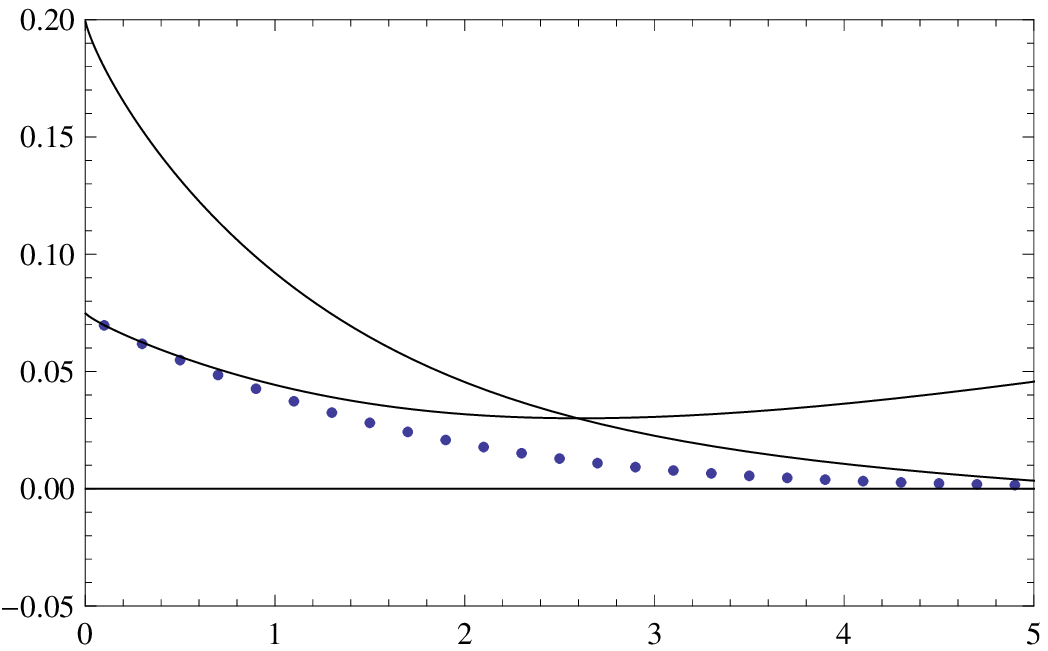}}
\label{fig:kmp1b}}
\hfill
\subfigure[The renormalised, rescaled data]{\scalebox{0.7}{\includegraphics{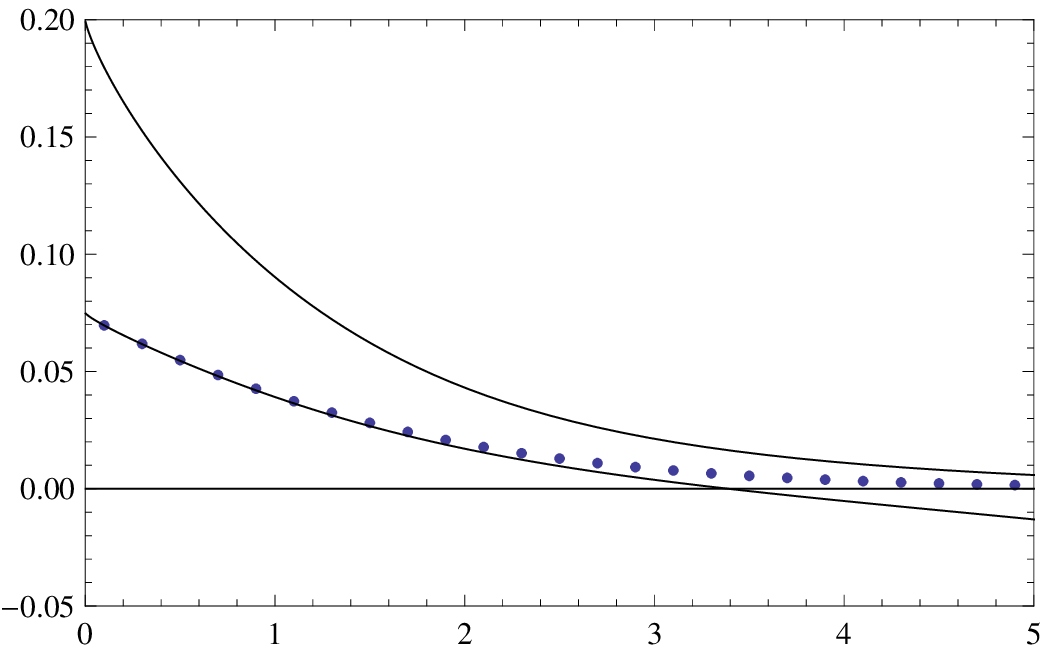}}
\label{fig:kmp2b}}
\caption{The scaling gaps for the massive perturbation $\Mc A_4^{(-)}$ at
truncation levels 5 together with the exact TBA results for the first
gap (dotted).
} 
\label{fig:kmp12b}
\end{figure}

Another noticeable effect of the rescaling is to improve the behaviour
of the ground states in each sector - the inclusion of the shifts
$\delta_i$ makes a qualitative difference in the convergence of the
ground states; we show these regions in figure \ref{fig:kmp12b} where
we plot the energy gaps from the 
bare TCSA data and the RG improved TCSA data together with the 
difference of the ground states in the $(1,*)$ and $(2,*)$
sector calculated using the TBA system of \cite{KMB370}.
There is a clear difference
between the ground states at $r=5$ whereas in figure \ref{fig:kmp2b}
this has been substantially decreased and the region of good agreement
with the TBA extended from the region $r<0.6$ to $r<2$.

We can show that it is only the form of the renormalisation and
re-scalings used in figure \ref{fig:kmp2} that are wrong by choosing to
fit the TCSA data to the BY data, and so deduce effective re-scalings
and renormalisations. We choose to fit the first two gaps in the even
sector to find these numerical renormalisation and re-scaling, and use
these for the rest of the data. This empirically renormalised and
rescaled TCSA data is shown in figure \ref{fig:kmp3} along with the BY
data, where we have chosen to remove the exponential corrections to
the ground states in each sector and set them to be zero. 
The agreement is impressive, even for $r$ as large as 12,
confirming that the TCSA data is indeed very accurate once the
renormalisation and rescaling has been taken into account.
To give some idea of the size, at $r=12$ the empirically calculated
rescaling and renormalisation functions at level 9 are
$r_{9}=0.76$ and $g_9 =0.71$. 

We can use the empirically calculated rescaling to check the scaling
form predicted from \eref{eq:rngndef}, that is 
\be
  r_n(\lambda) = r(\lambda n^{(1-2y)/2})
\;,\;\;
  r_{\text{1-loop}}(x) = 
\left(1 + \frac{x^2}{4\Gamma(2h)^2(2\pi)^{2y-1}} \right)
\;.
\ee
In figure \ref{fig:kmp4b},
Correspondingly, we plot the estimate of and 1-loop approximation to
$r(x)$ in figure \ref{fig:kmp4b}. We see that there is actually
good agreement with both the scaling form and numerical value of the
function.
%, and it is consistent with the 1-loop function being exact.

\begin{figure}[thb]
\centerline{\scalebox{0.7}{\includegraphics{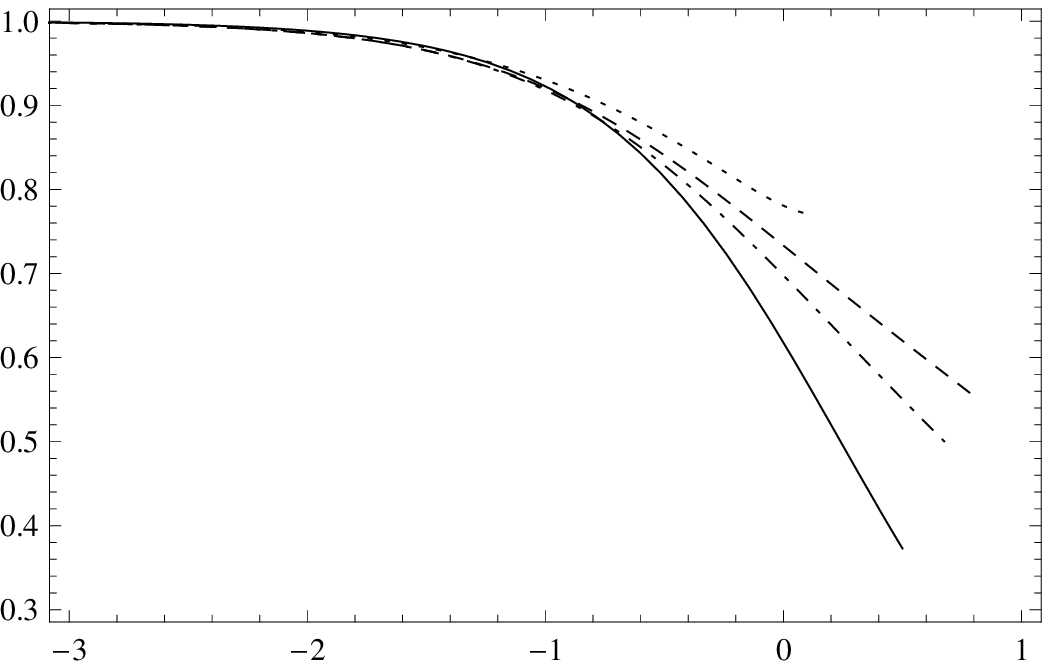}}}
\caption{The function $r(x)$ vs. $\log(x)$ from the fitting of the
odd sector to the BY lines at truncation levels 3 (dotted), 6
(dashed) and 9 (dot-dashed) as well as the perturbative
prediction $r_{\text{1-loop}}$ (solid)}
\label{fig:kmp4b}
\end{figure}

This can not be said of the renormalisation function $g_n(x)$. The
inaccuracy of the BY approximation for small $\lambda$ or $r$ means
that we cannot effectively use it to check the form of the
perturbative renormalisation function $g_n(x)$. For large values of
$r$ we see a definite disagreement with both the numerical value of and
the scaling form of the 1-loop prediction. For large values of $r$, we
see $g_n(x) \sim g( x n^{-0.3})$, the same form as the rescaling
function. Since this is a larger exponent than that predicted from
1-loop, which would be $g(x n^{-0.6})$, it is quite possible that the
larger exponent is correct, arising from higher loop corrections, just
as the exponent for the rescaling comes from a second order effect. 
The best hope we have of checking this in detail is for the exact TBA
excited state spectrum to be calculated on the cylinder which would
allow comparison with $g_{\text{1-loop}}$ down to small values of $r$.

\blank{
\begin{figure}[htb]
\subfigure[The function $r(x)$ vs. $\log(x)$ from the fitting of the
even sector to the BA lines at truncation levels 3 (dotted), 6
(dashed) and 9 (dot-dashed) as well as the perturbative
prediction (solid)]{\scalebox{0.7}{\includegraphics{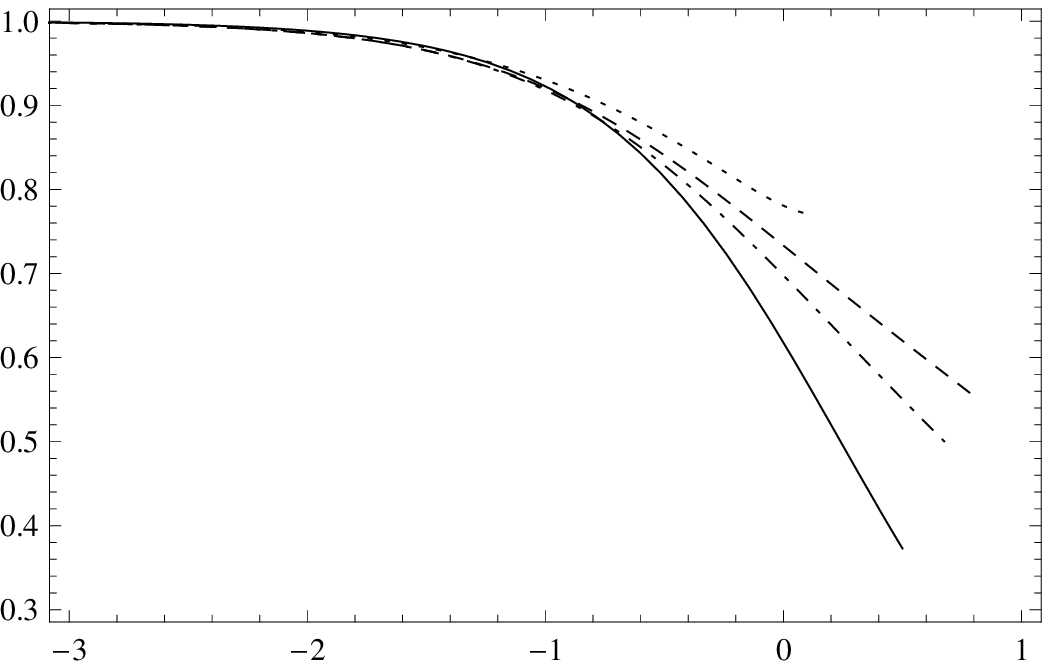}}
\label{fig:kmp4b}}
\caption{
\small The TCSA data fitted to the BA lines and the energy rescaling function
calculated from this fitting. } 
\label{fig:kmp34}
\end{figure}
}

\sect{Divergence for $h>3/4$}
\label{sec:div}

One of the main predictions of equation \eref{eq:rngndef} for the
rescaling functions is the exponent of $n^{4h-3} = n^{1-2y}$. This
goes to zero for large $n$ for $h<3/4$ but diverges for $h>3/4$, so
the rescaling function will diverge with increasing $n$ for $h>3/4$.
Since this is the leading $n$-behaviour in the scaling function gaps,
it suggests strongly that the TCSA scaling function gaps will only
converge to the exact answer for $h<3/4$. 
This means that any application of the bulk TCSA on the cylinder for
$h>3/4$ will only be able to predict the ratio of energy gaps.
As a demonstration, we give here the first few scaling function gaps
for $\Mc A_4^{(+)}$ with $h=3/5, 4h-3=-3/5$ and for $\Mc A_9^{(+)}$
with $h=4/5, 4h-3=1/5$. 
As can be seen, the gaps decrease with increasing level for 
$\Mc A_4^{(+)}$, in accordance with the prediction that the rescaling
function decreases with increasing $n$. On the other hand, the gaps
increase for $\Mc A_9^{(+)}$, showing no sign of convergence, in
accordance with the divergence of the perturbative rescaling function.

\begin{figure}[thb]
\subfigure[$\Mc A_4^{(+)}$]{\scalebox{0.406}{\includegraphics{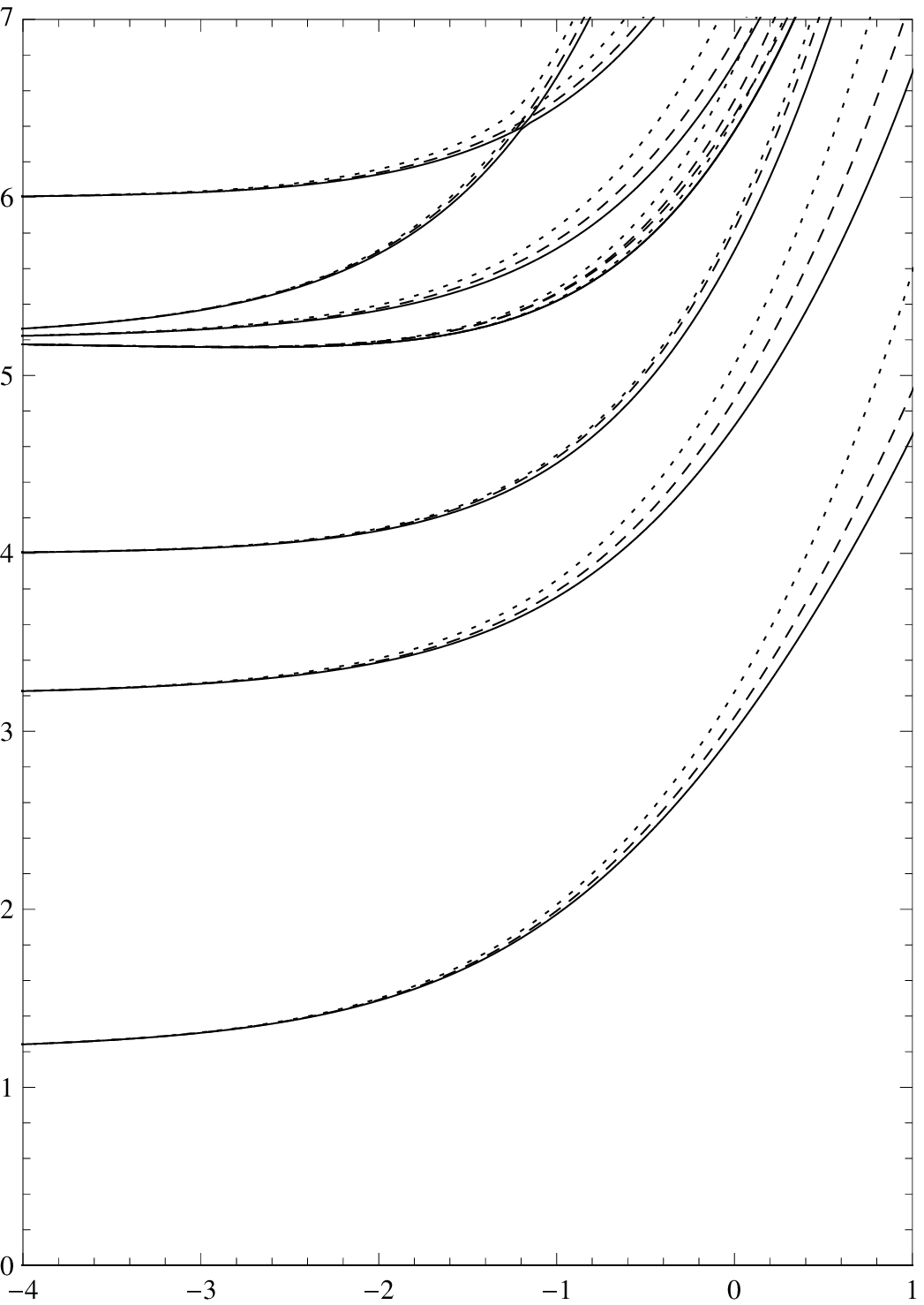}}
\label{fig:sp1}}
\hfill
\subfigure[$\Mc A_9^{(+)}$]{\scalebox{0.406}{\includegraphics{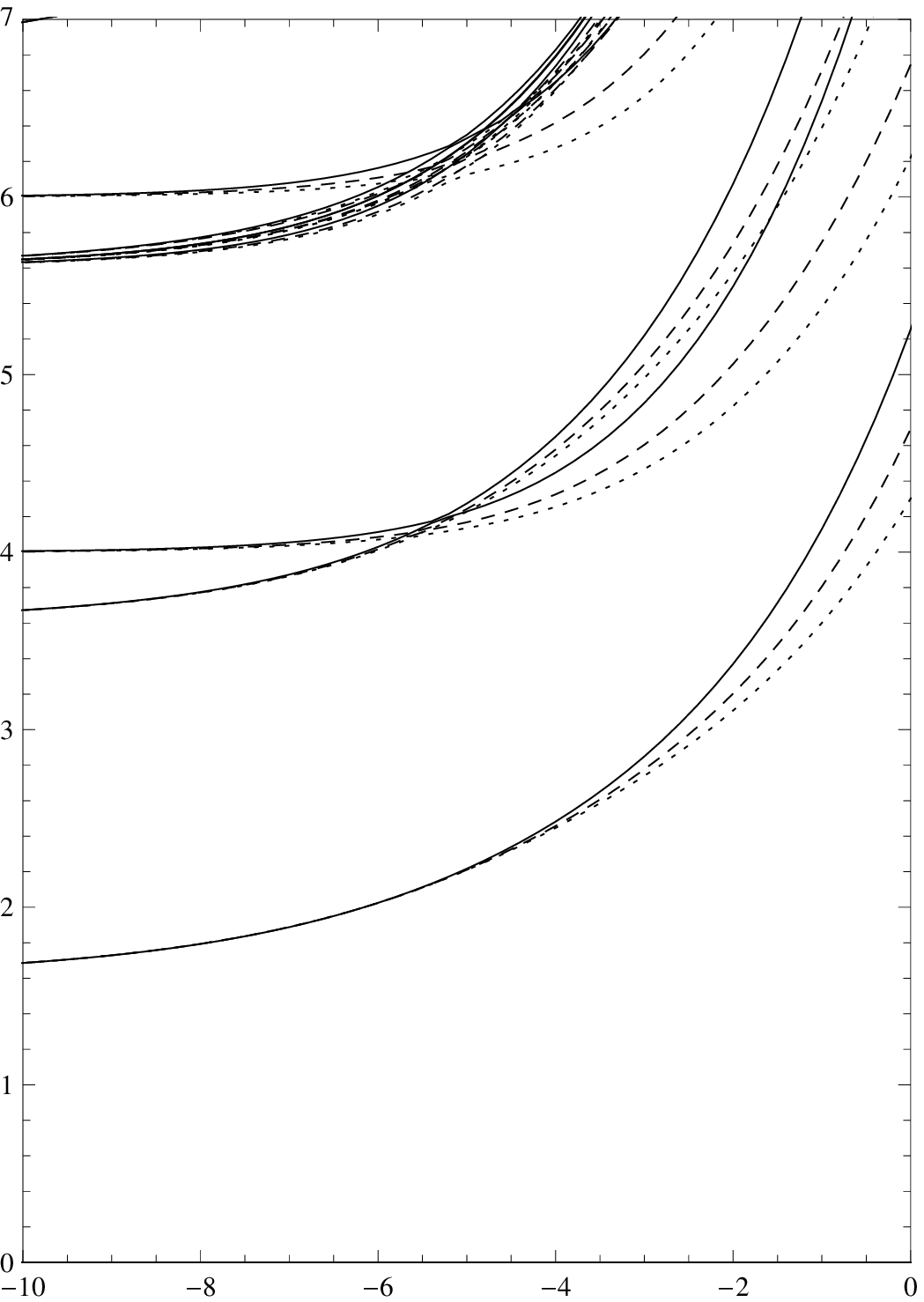}}
\label{fig:sp2}}
\hfill
\subfigure[$\Mc A_9^{(+)}$]{\scalebox{0.406}{\includegraphics{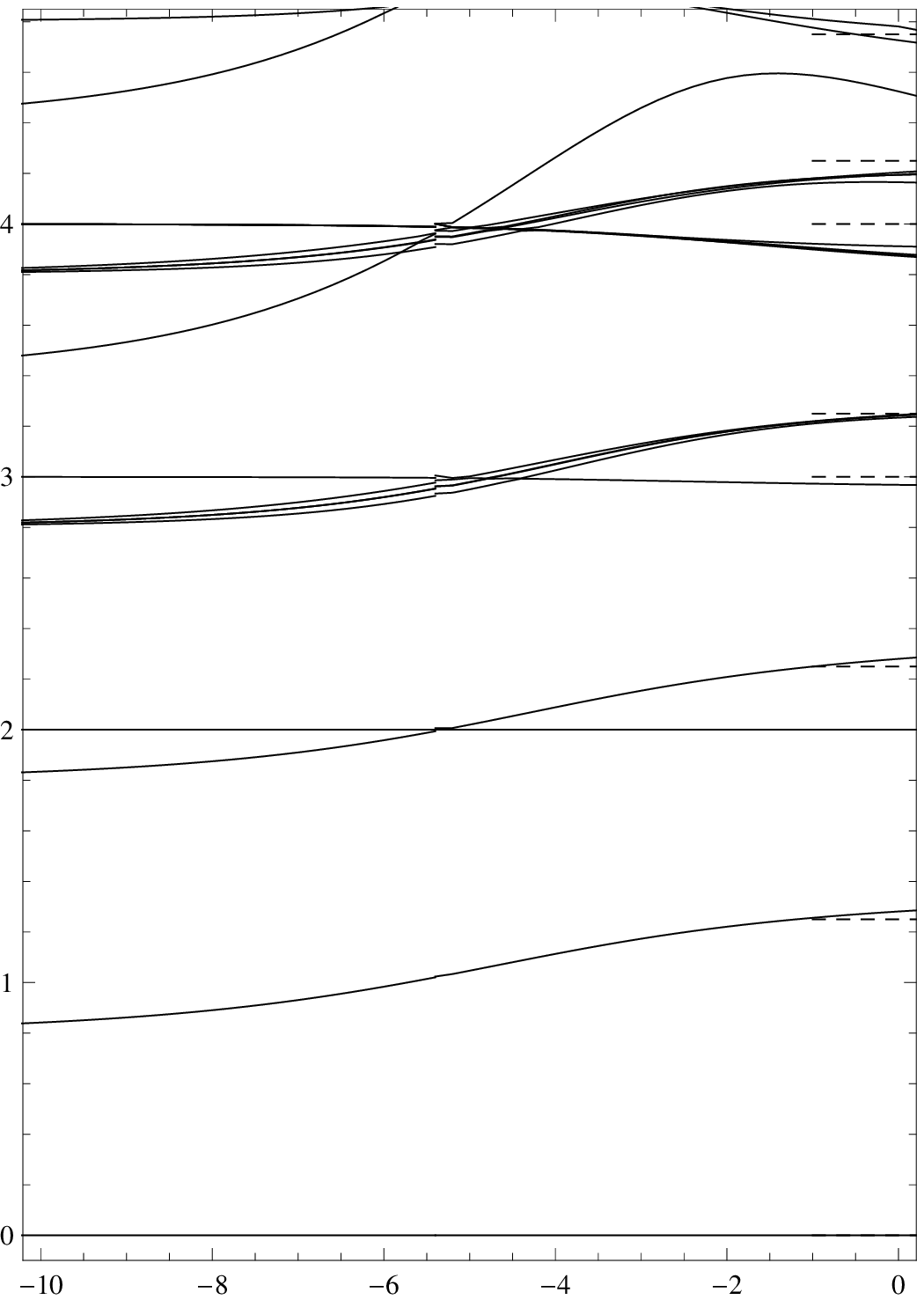}}
\label{fig:m910}}
\caption{\small The bare TCSA gaps for the $(1,*)$ sectors of (a) $\Mc
  A_4^{(+)}$ 
at truncation levels 3 (dotted), 6 (dashed) and 10
(solid), 
and (b) $\Mc A_9^{(+)}$
at levels 5 (dotted), 6 (dashed) and 10 (solid); (c) shows the ratios
of  gaps in $\Mc_9^{(+)}$ at level 10.
} 
\label{fig:sp12}
\end{figure}

This does not mean that the TCSA contains no useful data - there is
good evidence that the ratio of the energy gaps still remains
physical, even if the gaps themselves diverge. In figure
\ref{fig:m910} we show the ratio of the energy gaps  
$2 \tilde e_i/\tilde e_3$ for the same range
as \ref{fig:sp2}, showing that they are converging nicely to the
expected $(*,1)$ sector of $M_{8,9}$ (the dashed lines on the right of
that plot). 

\sect{Complications for bulk perturbations on the strip}
\label{sec:strip}

\begin{figure}[thb]
\subfigure[The normalised gaps for the $(1,*)$ sector of $\Mc A_4^{(-)}$ on the cylinder]{\scalebox{0.403}{\includegraphics{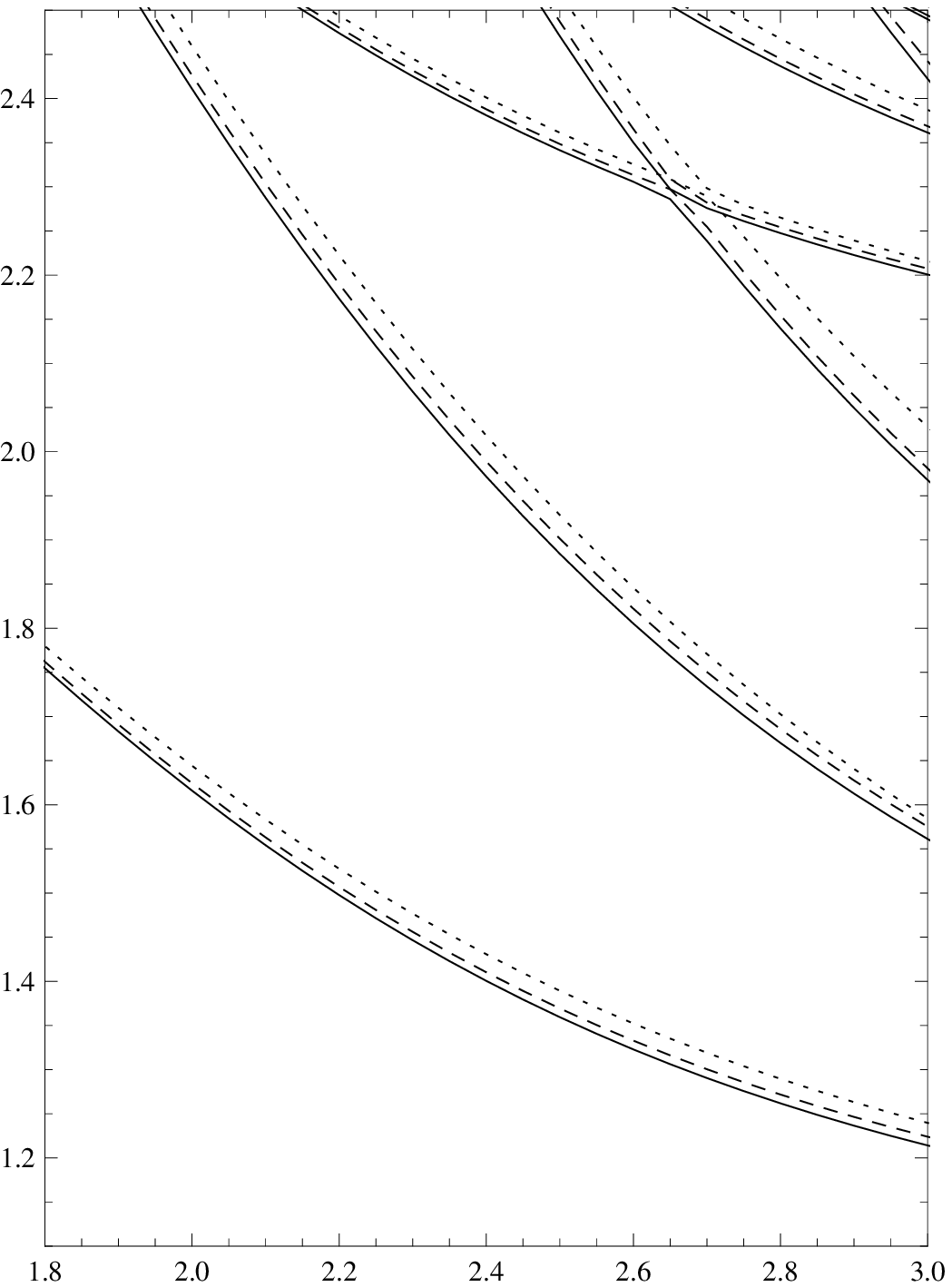}}
\label{fig:rg1}}
\hfill
\subfigure[The normalised gaps for the $(1,*)$ sector of $\Mc A_4^{(-)}$ on the cylinder with
renormalised coupling constant]{\scalebox{0.403}{\includegraphics{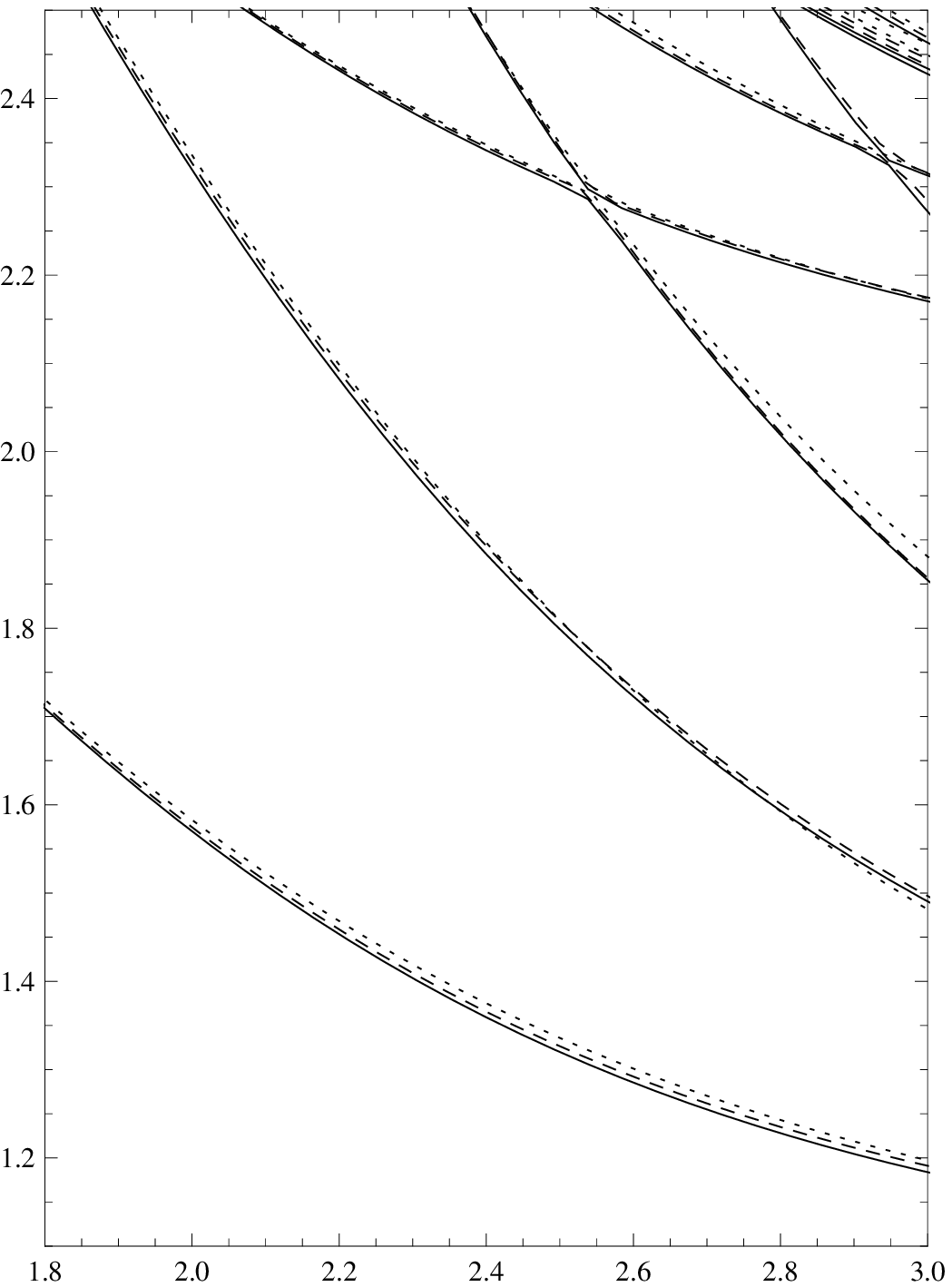}}
\label{fig:rg2}}
\hfill
\subfigure[The normalised gaps for the even sector of $\Mc A_4^{(-)}$ on the strip with fixed boundary conditions]{\scalebox{0.403}{\includegraphics{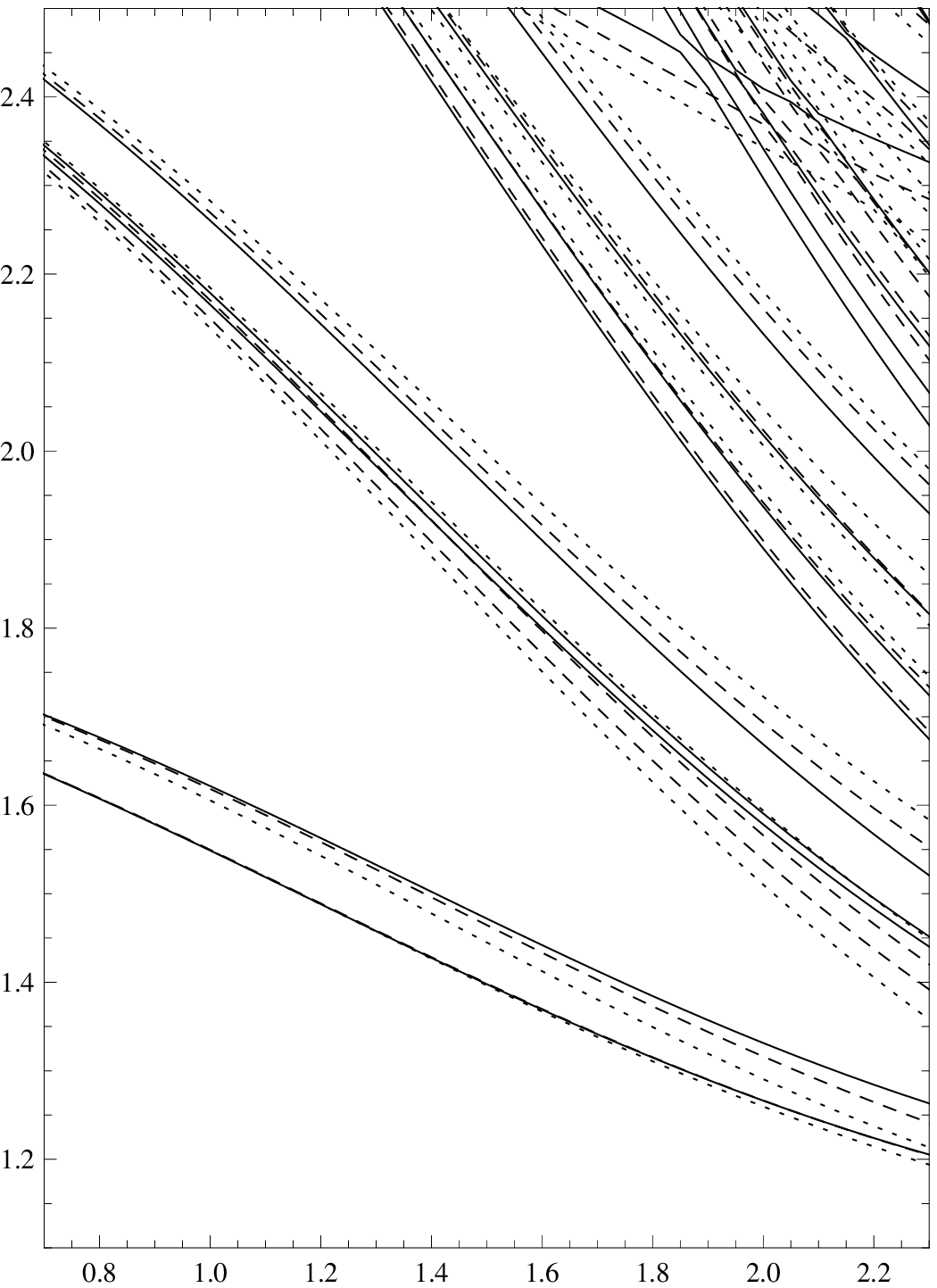}}
\label{fig:rg3}}
\caption{The model $\Mc A_4^{(-)}$: (a) the
  normalised gaps plotted vs. $r$ for truncation
  levels 6 (dotted), 8 (dashed) and 9 (solid) and
  (b) the 
  same gaps with renormalised $r$; (c) shows the model on the strip
  for truncation levels 10, 12 and 16 respectively.
} 
\label{fig:rg123}
\end{figure}

So far we have not discussed the TCSA method applied to bulk
perturbations on the strip. This was first used in \cite{Dorey:1997yg}
where it proved very accurate for the Lee-Yang model (for which
$y=12/5$).
We would like to make a similar analysis to that here to find the
leading truncation effects on the spectrum, but the form of the
correlation functions on the strip make this much harder and so far we
have not been able to find even the leading term such as a coupling
constant renormalisation. There is however, numerical evidence to
suggest that there are in fact no such simple renormalisation and
rescaling effects on the strip. In figure \ref{fig:rg123} we show the
effect of changing truncation level on the scaled energy gaps in the
massive perturbation $\Mc A_4^{(-)}$ on the cylinder and the strip. 
Since we look at the ratio of energy gaps, the ground state energy and
rescaling terms are removed, and we should see simply the effect of
the coupling constant renormalisation.
In figure \ref{fig:rg1} this is exactly what we see for the model on
the cylinder: as the level is increased, the lines all move to the
left, suggesting that a renormalisation of the coupling is
required. In figure \ref{fig:rg2} we perform the 1-loop coupling
constant renormalisation from \ref{eq:rngndef} and indeed the TCSA
data from the different levels are renormalised onto a single set of
lines. 

In figure \ref{fig:rg3}, however, we show the scaled energy gaps for
the same model on the strip with $(1,1)$ boundary conditions on each
edge. In this case, as the truncation level is altered there is no
consistent movement of the lines to the left or the right, indeed the
second gap appears to be almost invariant under the change of the
level. It is clear that no single coupling constant renormalisation
will make the TCSA data from the various levels map into a single set
of lines; the leading truncation effect on the TCSA on the strip is
not a simple coupling constant renormalisation.

\blank{
\sect{Irrelevant perturbations}
\label{sec:ir}
\begin{figure}[thb]
\centerline{\scalebox{0.7}{\includegraphics{IRplot1.eps}}}
\caption{The normalised gaps for the irrelevant perturbation $\Mc A_4
  + \varphi_{(3,1)}$ in the sector $(*,1)$ at truncation level 27,
  showing the flow to the $(1,*)$ sector of $\Mc A_5$.. }
\label{fig:ir1}
\end{figure}
}
\sect{Conclusions}

We believe we have presented good evidence that the leading
corrections to the TCSA method on a cylinder are a possible
ground-state divergence, a coupling constant
renormalisation and a representation-dependent energy re-scaling.
Application of these to the early tests of TCSA in \cite{LCM1} and
\cite{KM1} have greatly improved the accuracy and reliability of the
TCSA method which appeared to be very poor in those cases.
One by-product has been the important result that the TCSA does not
actually converge for perturbations with $h>3/4$ where there is a
divergent energy rescaling, but we have presented evidence that the
ratio of energy gaps does still converge in this case.

We would like to point out that the perturbative renormalisation and
rescaling use no more information than is usually available when using
the TCSA -- the scaling dimensions, three-point couplings and modular
S--matrix. In this way we think of them as an improvement of the TCSA.
They do not require knowledge of the four-point functions or conformal
blocks which can be difficult to calculate, even for minimal models
\cite{YBk}. 

It has also been suggested  \cite{Dorey} that the TCSA may have a
finite radius of 
convergence for any $n$. The apparent lack of convergence
in $n$ for the first even gap of $\Mc A_4^{(+)}$ in figure \ref{fig:cp4}
might be a sign of this, although since this occurs when the scaling
function is greater than its IR value, it is also possible as we
suggest that this divergence occurs after the IR fixed point is
reached, which we expect to be at a finite value of $n$ tending to
infinity as $n$ increases.

The results have found now allow one to apply the TCSA in a wide variety of
cases where it was not thought useful before, in particular we hope to
use it to study irrelevant perturbations in the light of the results
on irrelevant boundary perturbations in \cite{W2011}.

\sect{Acknowledgements}

%I am very grateful to P.A.~Pearce for
%their collaboration on \cite{FGPTW} and who
%providing the numerical data for the TBA flows from his papers
%\cite{PCA1,PCA2}, without which this paper would not have been possible.

PG would like to thanks STFC for support under studentship
PPA/S/S/2005/04104 and ST/I505748/1.
GMTW would like to thank G.~Tak\'acs and G.Zs.~T\'oth for helpful
discussions, G.~Tak\'acs for comments on the paper and STFC grant
ST/G000395/1 for support. 
All numerical calculations were performed using Mathematica \cite{mathematica}.

\label{sec:acc}

\newcommand\arxiv[2]      {\href{http://arXiv.org/abs/#1}{#2}}
\newcommand\doi[2]        {\href{http://dx.doi.org/#1}{#2}}
\newcommand\httpurl[2]    {\href{http://#1}{#2}}

% \newpage

\end{document}